\documentclass{agujournal2019}
\usepackage{url} 

\usepackage[utf8]{inputenc}
\usepackage[T1]{fontenc}
\usepackage[percent]{overpic}

\journalname{Journal of Advances in Modeling Earth Systems}

\begin{document}

\title{Machine learning emulation of gravity wave drag in numerical weather forecasting}

\authors{Matthew Chantry\affil{1}, Sam Hatfield\affil{2}, Peter Dueben\affil{2}, Inna Polichtchouk\affil{2}, Tim Palmer\affil{1}}
\affiliation{1}{Atmospheric, Oceanic and Planetary Physics, University of Oxford, Oxford, United Kingdom}
\affiliation{2}{European Centre for Medium-Range Weather Forecasts, Reading, United Kingdom}

\correspondingauthor{Matthew Chantry}{matthew.chantry@physics.ox.ac.uk}

\newcommand{\red}[1]{\textcolor{red}{#1}}

\begin{keypoints}
\item {Non-orographic gravity wave drag parametrisation can be accurately emulated with a neural network.}
\item These emulators produce accurate \& stable forecasts over long timescales.
\item {Neural networks can reduce the cost of an increased complexity scheme.}
\end{keypoints}

\begin{abstract}
We assess the value of machine learning as an accelerator for the parameterisation schemes of operational weather forecasting systems, specifically the parameterisation of non-orographic gravity wave drag. Emulators of this scheme can be trained {to} produce stable and accurate results up to seasonal forecasting timescales. Generally, more complex networks produce more accurate emulators. By training on an increased complexity version of the existing parameterisation scheme we build emulators that produce more accurate forecasts. {For medium range forecasting we find evidence our emulators are more accurate} than the {version of the parametrisation scheme that is used for operational predictions}. Using the current operational CPU hardware our emulators have a similar computational cost to the existing scheme, but are heavily limited by data movement. On GPU hardware our emulators perform ten times faster than the existing scheme on a CPU.
\end{abstract}

\section*{Plain Language Summary}
The ability for computers to construct models from data (machine learning) has had significant impacts on many areas of science. Here we use this ability to construct a model of an element of a numerical weather forecasting system. This element captures one physical process in the model, {a part of the model that describes the propagation of large-scale waves through the atmosphere}, but the long term aim would be to make many models each capturing a process. The goal is that the computer generated model will perform the task more efficiently than the existing model. Testing is then carried out to ensure that our computer model performs as well as the existing model. This is a challenging step, as learning is done over short time periods (seconds), but forecasts need to be accurate over years. Our computer generated models produce accurate forecasts on all tested timescales. On current computers they are not faster, but will be if weather forecasting centres invest in computers with graphics processing units.

\section{Introduction}

Numerical weather prediction has a proud history of saving lives and protecting property in societies particularly vulnerable to extremes of weather. As these extremes become more extreme still under the influence of climate change, it is important that numerical weather prediction systems improve further. One way to enhance the skill of numerical weather prediction is to increase model resolution. Not only will higher resolution directly enhance the ability of models to simulate small-scale extreme events, higher resolution will also enable the information in observations to be better assimilated at initial time, and will reduce the dependence of models on inaccurate parameterised processes, and in this way will reduce systematic errors \cite{palmer2020vision}.
However, increasing model resolution is computationally expensive: a doubling of resolution can increase computing costs by up to a factor of 16. Therefore we must find ways of improving the computational efficiency of our models, so that valuable computing resources can be targeted where they will be most effective.

Many physical processes that are involved in the forecasting of weather or climate occur at spatial scales smaller than the numerical grid of the models. Therefore while the physics might be understood, it is necessary to derive closure schemes which capture the effect of these physical processes on the grid scales. Examples of physical processes are radiation, convection and gravity wave drag, {the latter of }which will be our focus here. Inherently, by the nature of being a closure scheme, these physical parameterisation schemes are uncertain and imperfect. Together these schemes typically account for a significant portion of both the computational burden  and the number of lines of code in the code base. In the European Centre for Medium-Range Weather Forecast's (ECMWF) Integrated Forecasting System (IFS) model they contribute to about a third of the overall computational cost of running the model.
As more Earth System complexity is introduced into numerical weather prediction models, e.g. aerosols and atmospheric chemistry, these numbers will only increase.

The recent boom in both hardware and software developments around machine learning has caused many fields to examine the possible boons that machine learning can bring. In the field of weather and climate forecasting researchers are examining the applicability of machine learning techniques to a spectrum of problems \cite{chantryMLintro}, covering changes from the {seismic} to the incremental. {Seismic changes include investigations into} whether machine learning can replace the whole forecasting system, either by learning from observational data \cite{metnet} or atmospheric reanalysis \cite{weatherbench}. Early results are promising in the area of nowcasting \cite{metnet}, but still lag behind classical modelling for short and medium range forecasting \cite{weyn2019can} {with evidence \cite{rasp2021data} and arguments \cite{palmer2020vision} that there is insufficient data when moving to high resolutions.}
For seasonal forecasting machine learning techniques again show promising results, e.g. forecasting El Nino sea-surface temperatures \cite{dijkstra2019application}.
At the {incremental end of the} spectrum lie approaches where machine learning aims to learn the behaviour of existing kernels of a weather forecasting model, with the aim of building machine learning emulators that can accelerate the kernel \cite{krasnopolsky1997neural,chevallier1998neural,o2018using,brenowitz2018prognostic}. Successful acceleration of these kernels could allow reductions in run-time or re-investments of the computation savings into increased complexity kernels or increases in spatial resolution.  It is this approach that we shall investigate here. The appeal of this approach is the ability to leverage existing physics knowledge in a small self-contained unit. This application of machine learning to weather and climate forecasting is closely related to the use of reduced numerical precision to accelerate weather forecasting \cite{vavna2017single,hatfield2019accelerating}, whereby a slightly less accurate version of a kernel can be used undetected beneath the uncertainty and inaccuracy of the system. On current GPUs neural networks can be easily run at reduced precision for both training and inference, leveraging increased linear algebra performance \cite{nvidia2017}. While existing atmospheric algorithms need to be reformulated to fit within the compact dynamic range of half-precision \cite{hatfield2019accelerating}, neural networks naturally fit within this range due to data normalisation.

Several groups have already {tackled} the wider problem of physical parameterisation schemes. {These efforts fall into two groups, one seeks to build new parametrisation schemes by training on observations or higher-resolution models, e.g. \citeA{brenowitz2019spatially} who built a neural network to learn all parameterised physics by coarse-graining higher-resolution aqua-planet simulations. The second aims to emulate an existing parametrisation scheme in order to increase the performance.} This approach started in the 1990s when both \citeA{krasnopolsky1997neural} and \citeA{chevallier1998neural} used neural networks to emulate and accelerate radiation schemes. Their approaches were broadly successful but did not lead to widespread adoption within operational weather or climate models. {In the case of \citeA{chevallier1998neural}, when the number of vertical layers increased beyond 60 layers, the emulators could not be trained to work at sufficient accuracy \citeA{morcrette2008reduced}.} More recently \citeA{yuval2020stable} used random forests to emulate a convection scheme. Efforts have also been made to accelerate increased complexity (dubbed super) parameterisation schemes \cite{rasp2018deep,gentine2018could}. \citeA{venmoRad} emulate a radiation scheme and assess the computational cost relative to the existing scheme. {\citeA{ukkonen2020accelerating} emulate the gas optics scheme within a radiation scheme, providing acceleration for this kernel.} \citeA{gettelman2020machine} use neural networks to learn an increased complexity microphysics model and reduce the cost down to that of their reference scheme.
{Of this recent body of work, most have involved some simplification steps relative to an operational forecast model. These include reduced horizontal or vertical grids and aqua-planet configurations. Here we will assess our emulators when coupled to models {running on the 25km grid used for long-range forecasting at ECMWF,} a significant step towards operational forecasting.}

Gravity wave drag is a parameterised physical process that has not yet been considered for emulation. It is typically parameterised as two processes, orographic and non-orographic. {In the IFS model these are two separate schemes} and it is the latter that we shall mostly focus on in this study. {Orographic gravity wave drag is discussed in section \ref{sec:Discussion}}. Non-orographic gravity waves can be generated by dynamics such as fronts and convection and occur on a vast range of scales \cite{gardner1989rayleigh,ern2004absolute} meaning that while some gravity waves are resolved by current forecast resolutions, others need to be parameterised. 
In the 
IFS model this is achieved using the scheme from \citeA{warner2001ultrasimple} and in particular the variant described in \citeA{orr2010improved} (henceforth referred to as NOGWD). Like most current physical parameterisation schemes, it operates on a vertical column of fluid and can be run in parallel across all the columns within a given grid. In this scheme a fixed amount of momentum is launched upwards from a given height at a range of {launch} angles in the horizontal direction and over a range of {phase speeds}. At each model level the stability of these waves within the atmospheric profile is calculated, and depending on this stability momentum is deposited at these layers. Accurate parameterisation of non-orographic gravity wave drag is important for maintaining an accurate zonal-mean wind and temperature distribution \cite{garcia1994downward} and for capturing the phase and amplitude of the Quasi-Biennial Oscillation (QBO) \cite{dunkerton1997role}. 
We choose to emulate NOGWD as it is a medium complexity scheme that has particularly important effects on seasonal timescales. 
The effects can be seen both in the stratosphere and troposphere \cite{polichtchouk2018sensitivity,polichtchouk2018impact}.
Recent work has shown the challenges in using offline trained neural networks in free-running simulations \cite{brenowitz2020machine} and we will assess whether these problems also exist in our scenario.
By sitting in the middle of the complexity spectrum, NOGWD has enough complexity to challenge machine learning methods but it is nontrivial that an emulator will be faster than the current scheme. If we can demonstrate efficiency gains in this parameterisation scheme, we would postulate that many of the parameterisation schemes can be efficiently emulated using neural networks.

Next, in section \ref{sec:data}, we discuss the data used for the machine learning, followed by our machine learning methodology (section \ref{sec:ml}). {In section \ref{sec:res} we emulate the NOGWD scheme and test the performance both decoupled and coupled to the IFS model.} 
In section \ref{sec:Discussion} we address several important questions in taking parameterisation emulators into operational prediction systems. We also present new results and discussions on the key issues of conservation properties, computational cost, alternate network structures, generalisability and emulating other parameteristion schemes. 

\section{Data}
\label{sec:data}
\begin{figure}
    \centering
    \includegraphics[width=1.\columnwidth]{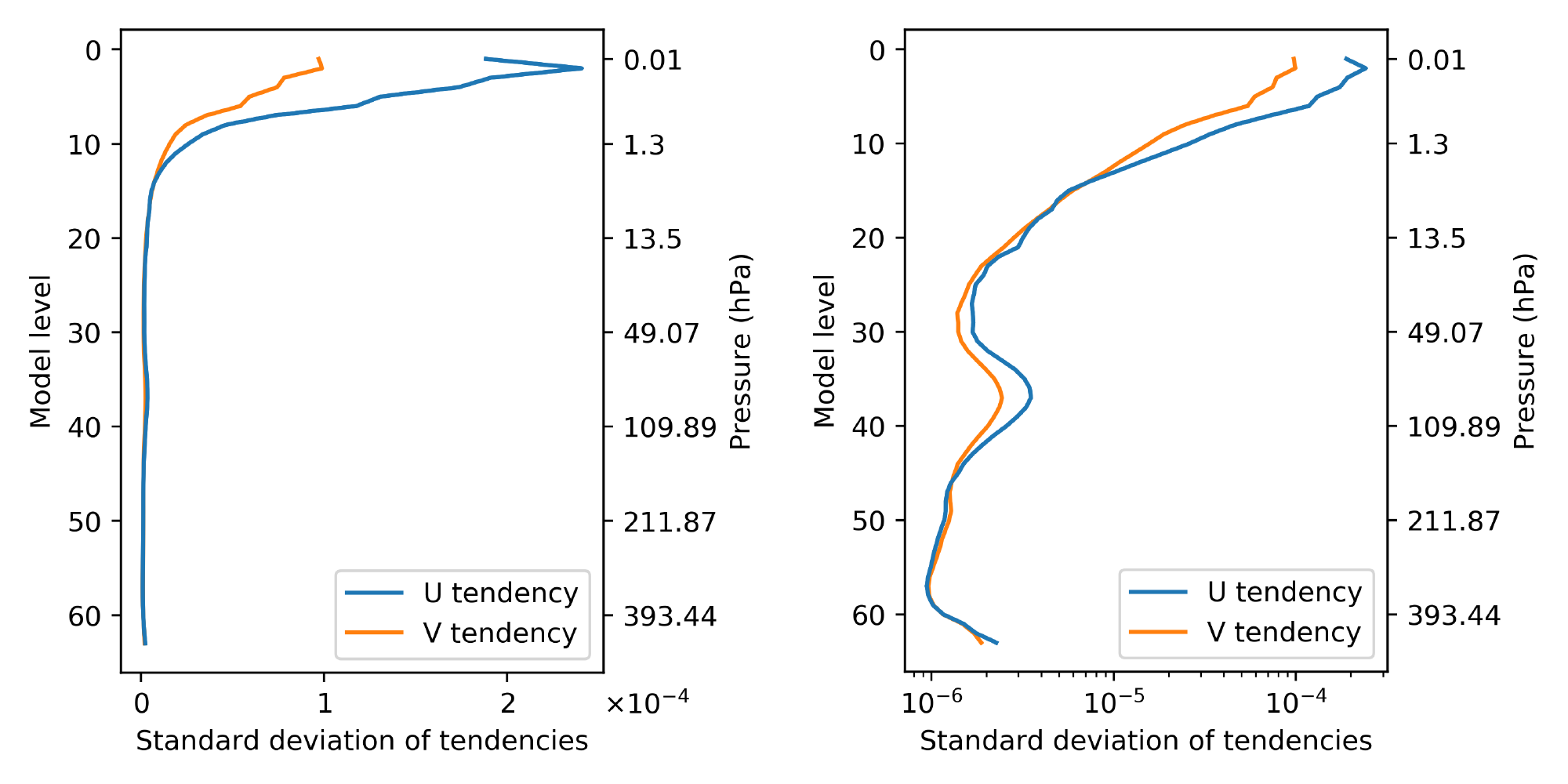}
    \caption{Standard deviation of the $u$ and $v$ wind tendencies from the HC NOGWD scheme (using the first 30 days of the dataset), plotted on regular (left) and semi-logarithmic (right) axes as a function of model level, plotting on the model levels where the scheme is activate. Velocity tendencies are largest at the top of the atmosphere (${\sim}0.01 \mathrm{hPa}$), with a second peak around the 30th model level (${\sim}70 \mathrm{hPa}$).
    }
    \label{fig:tendencies}
\end{figure}
The specific goal of this project is to create emulators of the non-orographic gravity wave drag parameterisation found in the IFS model. To generate the data we modify cycle 45R1 of IFS {\cite<the operational cycle in 2018,>{ifscy45r1}} to output all time-dependent variables either inputted to or outputted from the scheme in question, saved on the native cubic-octahedral grid. We use the horizontal grid TCo399, equivalent to ${\sim}25\textrm{km}$ grid-point spacing. In the vertical dimension we use 91 model levels spanning the surface to $1\mathrm{Pa}$. 
As outlined above, momentum is launched at a set number of {launch angles in the horizontal plane}, and for a range of {phase speeds}. Typically four launch angles are used, {describing momentum in the four cardinal directions}. The wavespeed spectrum is discretised into 20 elements. {Further explanation of these parameters can be found in \cite{orr2010improved}.} When generating datasets, we increase these values to 16 and 100 respectively, which we refer to from hereon as the HC (high complexity) scheme. These changes increase the cost of the existing scheme by a factor of 20, but provide a higher resolution discretisation of the equations governing NOGWD. \citeA{scinocca2003accurate} suggest that 35 elements is sufficient but differences can still be observed between 35 and 1000 elements. 
{We observe that the HC variant of the NOGWD scheme produces more accurate forecasts (depicted later in figures \ref{fig:IVER_T} \& \ref{fig:IVER_W}).} 
In preliminary testing we found that our networks produce a lower mean-squared-error against the testing dataset when both training and testing data came from the HC scheme, compared with when the standard complexity scheme provides the testing and training data. This is perhaps because the HC has smoother vertical tendency profiles. In our online testing later we will compare both the standard formulation (with 4 and 20) and the HC variant of the existing scheme.

Here, IFS is run for 30-day integrations, {re}starting every 30 days, {in total} covering a three year period. This was chosen to give a full seasonal cycle for each of training, testing and validation data. Data is saved every 20 hours, in this way sampling the daily cycle. {At our chosen resolution, there are $654400$ columns within our grid.} Without further reductions, each 30-day period produces $23558400$ input/output columns for training. We sub-sample every 10th  column, changing the starting index for different 30-day periods {(e.g. grid-columns 1, 11, 21... in the first period and 2, 12, 22... in the second period)}. In this fashion, every grid-point features in the training dataset. Our training dataset comprises of 12 30-day periods, covering 2015, totalling $28,270,080$ training pairs. Our {validation} and {test} datasets comprise of three 30-day periods spread across 2016 and 2017 respectively, (starting 2016-02-25, 2016-06-24, 2016-12-21 and 2017-02-19, 2017-07-19, 2017-11-16 respectively).  Our training dataset is significantly larger than is typically used in emulation training, e.g. \citeA{brenowitz2020machine} used $1797120$ columns in training, 
{although notably smaller than the 140 million examples used in \citeA{rasp2018deep}}. We tested reducing the data volume by training only on the first $10\%$ of each month, but training for 10-times the number of epochs {(number of complete passes through the data during training)}. Training with this data gave a significant degradation in our testing error (${\sim}40\%$ larger), {but did not show any obvious signs of over-fitting}. {For training, using a large dataset but iterating for fewer epochs lead to better results on the validation and testing datasets. The size or precise period selected for the validation and testing dataset did not change the results. The increased training dataset provides increased generalisation ability on unseen data. 
}

To mirror the existing parameterisation scheme, we include only the variables used to predict the velocity tendencies in the existing model. These are the horizontal velocity ($u$ and $v$) and temperature profiles on the vertical model levels. In addition there are three descriptors of the model levels at each grid-point: the pressure, half-level pressure and geopotential. The outputs of the existing scheme are horizontal velocity tendencies, showing the impact of unresolved gravity wave drag on the velocity field. A temperature tendency can then be derived from the velocity fields and NOGWD tendencies using a simple formula\footnote{$T_t = - \frac{1}{c_p}\left( u u_t + v v_t \right)$, where $_t$ denotes the NOGWD tendency and $c_p$ is the dry air calorific capacity at constant pressure. }. In total, this produces a dataset with $6 \times 91 = 546$ inputs and $2 \times 91 = 182$ outputs.

Reductions can be made to both input and output vector sizes by applying simple domain knowledge. Firstly, examination of the existing scheme shows that the scheme is only active in the upper 63 layers of the atmosphere and only uses information in the top 63 layers of the input profiles. Secondly, pressure and half-level pressure can each be described on model levels as time-independent functions of surface pressure.
Theoretically, geopotential can be reconstructed using the surface geopotential, pressure, temperature and humidity, the last of which is not inputted to the scheme. In practice, we find our models produce accurate results using only surface pressure and surface geopotential.
Combining these ideas we produce input and output vectors of size $3 \times 63 + 2 = 191$ and $2 \times 63 = 126$ respectively. 
Even after data reduction the size of training data totals in excess of 30Gb, larger than {our CPU memory}. Therefore we re-write the data into the TFRecord format \cite{tensorflow} to allow easy and efficient streaming of data from disk to GPU. 
Data is publicly available in 30-day chunks\footnote{\url{https://storage.ecmwf.europeanweather.cloud/NOGWD_L91_PUBLIC/README.html}} in the HDF5 file format for portability.

Data normalisation is a typical step in any machine learning workflow. Here we normalise both the input and output data from our dataset. We examine two normalisation approaches. The first, here dubbed ``MEANSD'' , calculates elemental means and standard deviations for each feature (i.e. a variable at a given model level) and normalises both inputs and outputs by these values. The second, dubbed ``TMEANVMAX'', normalises temperature (plus pressure and geopotential) with the above method, but for each of velocity inputs and tendency outputs the entire column is divided by the largest standard deviation from the column and no mean is subtracted. 
With both methods our loss function will be the L2 loss (mean squared error). 
With the first normalisation, our models will seek to optimise each output feature equally, irrespective of the typical magnitude of the velocity tendency. 
With the second approach the model will be encouraged to learn features in proportion to their absolute size. The latter method might seem a more natural choice given that the outputs will contribute to the next velocity field values. However, if we examine the structure of the velocity profiles in figure \ref{fig:tendencies} we see that the largest tendencies occur at the very top of the model atmosphere. The top of the atmosphere is weakly constrained by data assimilation and velocity fields in the top ten layers of the 91 level IFS model are strongly damped with a sponge layer to prevent wave reflection from the rigid upper boundary. Therefore, allowing the model to focus on these layers might not be constructive for an accurate forecast when coupled back to the full atmospheric model. We will therefore try both approaches and later test both in coupled simulations.

\section{Machine learning methodology}
\label{sec:ml}
In this work we focus on neural networks (NN) as our tools for emulation. 
{For those unfamiliar with NNs we recommend \citeA{brenowitz2018prognostic} who describe NN in the context of emulation.}
Currently both neural networks and random forests are popular methodologies for parameterisation scheme emulation or creation. 
Random forests can conserve physical quantities  \cite<e.g. energy conservation in >{o2018using}, but recent work has shown how to also achieve this with neural networks \cite{beucler2019enforcing}.
Neural networks promise greater theoretical performance as random forests are limited by memory performance. 
However random forest methods have recently been shown in a cloud parameterisation scheme to be more stable in long-term simulations coupled to atmospheric models \cite{brenowitz2020machine}. With this in mind we will carefully assess whether our neural network models run stably when coupled to the IFS model for seasonal timescales to further investigate this issue. We use Tensorflow \cite{tensorflow} to build and train our networks.

Predominantly we will focus on fully-connected neural networks in our search space. These networks are the most general purpose, with no explicit encoding of the vertical structure of our dataset which would occur with a convolutional-based network. However, we will present results for more sophisticated network architectures in section \ref{sec:Discussion}.

Within the fully-connected neural network framework we wish to explore many of the remaining hyper-parameters. We constrain our networks to have constant width (for each layer) but explore the space of layer width (between 1 and 1000 neurons) and number of hidden layers (between 1 and 10 layers). For the activation function we test ReLU, tanh, leaky ReLU and Swish \cite{swish} activation functions, and consistently find that tanh and Swish produce the lowest training losses. After preliminary exploration we settle on using the Adam optimiser with a learning rate of $10^{-4}$ and a batch size of 256. We train for 50 epochs, check-pointing after each epoch to select the best model on the {validation} dataset.

\section{Results}
\label{sec:res}
Given our desire to accelerate an existing parameterisation scheme, the speed of any machine learning emulator built is as important as the accuracy. Therefore we present our offline results as a function of the degrees of freedom, DOF, (number of trainable parameters) in each network. For fully-connected neural networks this is a good proxy for the number of floating point operations (or FLOPs) required to calculate a single inference step on a column of data. The degrees of freedom in a fully-connected neural network are dominated by the neuron weights, each of which is used once in a fused multiply-add operation as part of a matrix-vector multiplication. FLOPs are only one way of measuring computational cost and are the appropriate measure when a calculation is operation-bound rather than memory-bound. However, as most parts of weather and climate models are not able to operate anywhere close to the peak performance of the hardware and as the ability to leverage peak performance will depend on pattern and data requirement of kernels, the measure of FLOPs may be misleading when estimating computing time. We will later return to this issue during online testing.

\subsection*{Offline learning}

\begin{figure}
    \includegraphics[width=0.95\columnwidth]{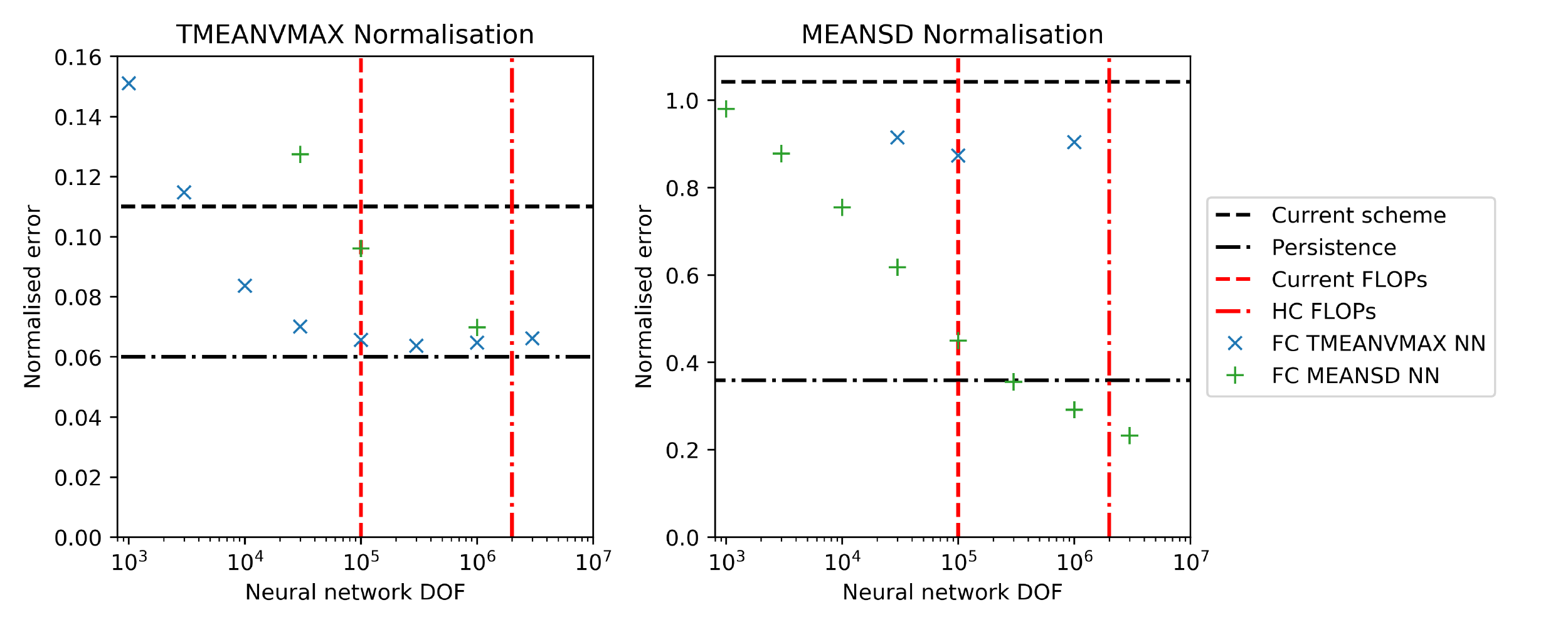}
    \caption{Offline root-mean-squared error of the best performing network for a range of prescribed DOFs. {Networks are trained and tested on data using two normalisation methods, dubbed MEANSD and TMEANVMAX. For each normalisation plot we plot models that were trained using both normalisation methods (not all models are plotted on each plot).} To contextualise the offline results we plot two baselines. First, the error of the current scheme relative to the HC scheme (which was used to generate the training data). Second, the error when persisting the HC scheme tendencies for the following timestep. A vertical red line denotes the approximate FLOP cost of the current scheme. 
    }
    \label{fig:offline}
\end{figure}

In figure \ref{fig:offline} we plot the performance of our best models for a given number of DOF on the test dataset for the two normalisation schemes. To give context to the error and cost of these schemes we plot the error on the test dataset of using either the existing scheme or persisting the HC scheme tendencies for the following timestep\footnote{In the operational version of IFS, persistence is used {every second hour} for the NOGWD scheme to reduce costs. Persistence on longer timescales would lead to a dramatically worse model.}.  Relative to the HC scheme, we are easily able to construct NN that produce lower error than the existing scheme. Beating persistence is a more challenging task that is only achieved for our most expensive MEANSD normalisation networks. For the largest TMEANVMAX networks our error no longer decreases and indeed shows a slight increase for the largest networks that we train. Potentially the addition of regularisation (e.g. on the weights) could help with training large networks.
{To understand the impact of normalisation we test a subset of networks trained with MEANSD by evaluating their outputs with the TMEANVMAX normalisation. For large networks we see similarly good scores as those trained with TMEANVMAX. Assessing TMEANVMAX networks using the MEANSD normalisation we see much poorer performance, even for large network sizes, suggesting that the smaller magnitude features have not been learnt as well. Coupled testing is required to explore whether these features have an impact on forecast quality.}
This offline testing phase helps identify the best network architectures for a given number of DOF. 
For both normalisation methods there is no clear pattern relating the depth, width and the DOF. {In table \ref{tab:models} we present our optimal networks which will be used in the coupled testing. For a given DOF choice we use our best network with the Tanh activation function as this has comparable performance to the Swish function but this activation function already exists with the Fortran standard.} These networks consist of 4 to 6 hidden layers, but test dataset losses vary little between 3 and 10 layers. This suggests that the precise network architecture is unimportant for these fully-connected fixed-width networks in our problem.

\begin{table}[h]
\begin{center}
\begin{tabular}{ |l|r|r|r|l| } 
 \hline
 Normalisation & DOF & Hidden layers & Hidden width & Nonlinearity \\ 
 \hline
 TMEANVMAX & $30,000$ & 51 & 6 & Tanh \\ 
 MEANSD & $30,000$ & 55 & 5 & Tanh \\ 
 TMEANVMAX & $100,000$ & 136 & 4 & Tanh \\ 
 MEANSD & $100,000$ & 122 & 5 & Tanh \\ 
 TMEANVMAX & $1,000,000$ & 461 & 5 & Tanh \\ 
 MEANSD & $1,000,000$ & 416 & 6 & Tanh \\ 
 \hline
\end{tabular}
\end{center}
\caption{Table of model architectures used in the coupled simulations.}
\label{tab:models}
\end{table}

\subsection*{Coupling neural networks within IFS}

Small offline errors are not a necessary or sufficient condition to find good performance when using the networks online with simulations with the IFS. Theoretically, seemingly small biases could accumulate as large errors and trigger instabilities even if the mean errors are small. 
To understand the impact of our schemes on the wider forecast we couple our neural networks back into the IFS code replacing the existing NOGWD scheme. This requires interfacing neural networks, typically written in Python, with Fortran, and is a problem faced by many machine learning researchers in this field. In our work we solve this problem by writing a Fortran module to load the weights saved and reproduce the neural network architecture using matrix-matrix multiplication algorithms found in the BLAS library \cite{blas}. This works well for simple network architectures but would be more difficult to realise for the complex network structures that can be built in libraries such as Tensorflow. Recently \citeA{fortrankeras} have produced an open-source package which accomplishes a similar task. 

\subsection*{Medium range forecasting}

\begin{figure}
    \centering
    \includegraphics[width=1.0\columnwidth]{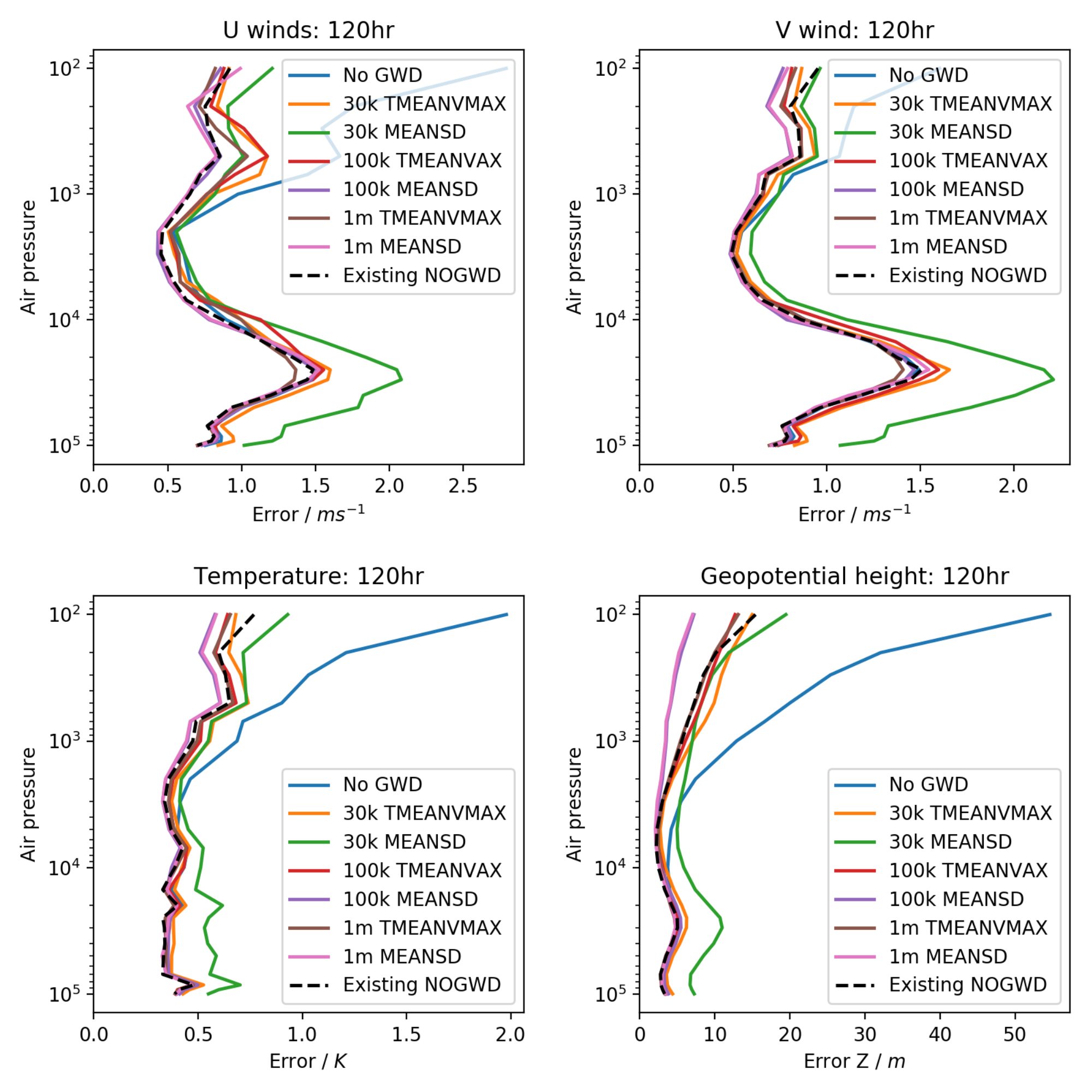}
    \caption{Horizontal-average forecast errors in simulations after 120 hours at TCo399 (${\sim}25 \mathrm{km}$) resolution. Truth is taken to be a simulation of IFS using the HC version of the NOGWD scheme. Against this we measure the existing NOGWD scheme, no NOGWD scheme and six variants of our NNs covering a range of complexities for both normalisation methods. Except for the $30,000$ (30k) DOF MEANSD NN each of the NNs have comparable errors to the original scheme, with several of the MEANSD schemes outperforming the original. }
    \label{fig:online}
\end{figure}

To demonstrate the performance of our networks in online mode, i.e. coupled to the IFS, we complete a 120 hour forecast at ${\sim}25 \mathrm{km}$ (TCo399) resolution using a variety of model configurations. Our truth here is chosen to be a simulation using the high-complexity variant of the existing NOGWD scheme. 
We choose the HC NOGWD scheme as truth instead of reanalysis because the aim is to emulate the HC scheme with a NN and not the atmospheric reanalysis.
Against this we plot the forecast error using the existing NOGWD scheme, using no activated NOGWD scheme at all and using six neural network configurations covering both normalisation methods and a range of network sizes. For each simulation we calculate the horizontally-averaged RMSE relative to a forecast using the HC scheme and plot the error as a function of pressure after 120 hours in figure \ref{fig:online} {(comparable results were found for other lead times)}. For the TMEANVMAX normalisation we find similar performance for each network, irrespective of network complexity, with each of the schemes producing a comparable forecast to the existing variant of the scheme. For the MEANSD normalisation, the higher complexity networks with $10^5$ and $10^6$ DOF produce the closest forecast to the truth, marginally more accurate than the existing version of the scheme. However the small MEANSD network produces a notably degraded forecast. This suggests that if one is trying to produce the cheapest possible forecast, then the TMEANVMAX normalisation allows a network to cheaply learn the key features. However to produce a best possible forecast there is value in a more equal weighting of the features, as given by the MEANSD normalisation.

\begin{figure}
    \centering
    \includegraphics[width=0.7\columnwidth]{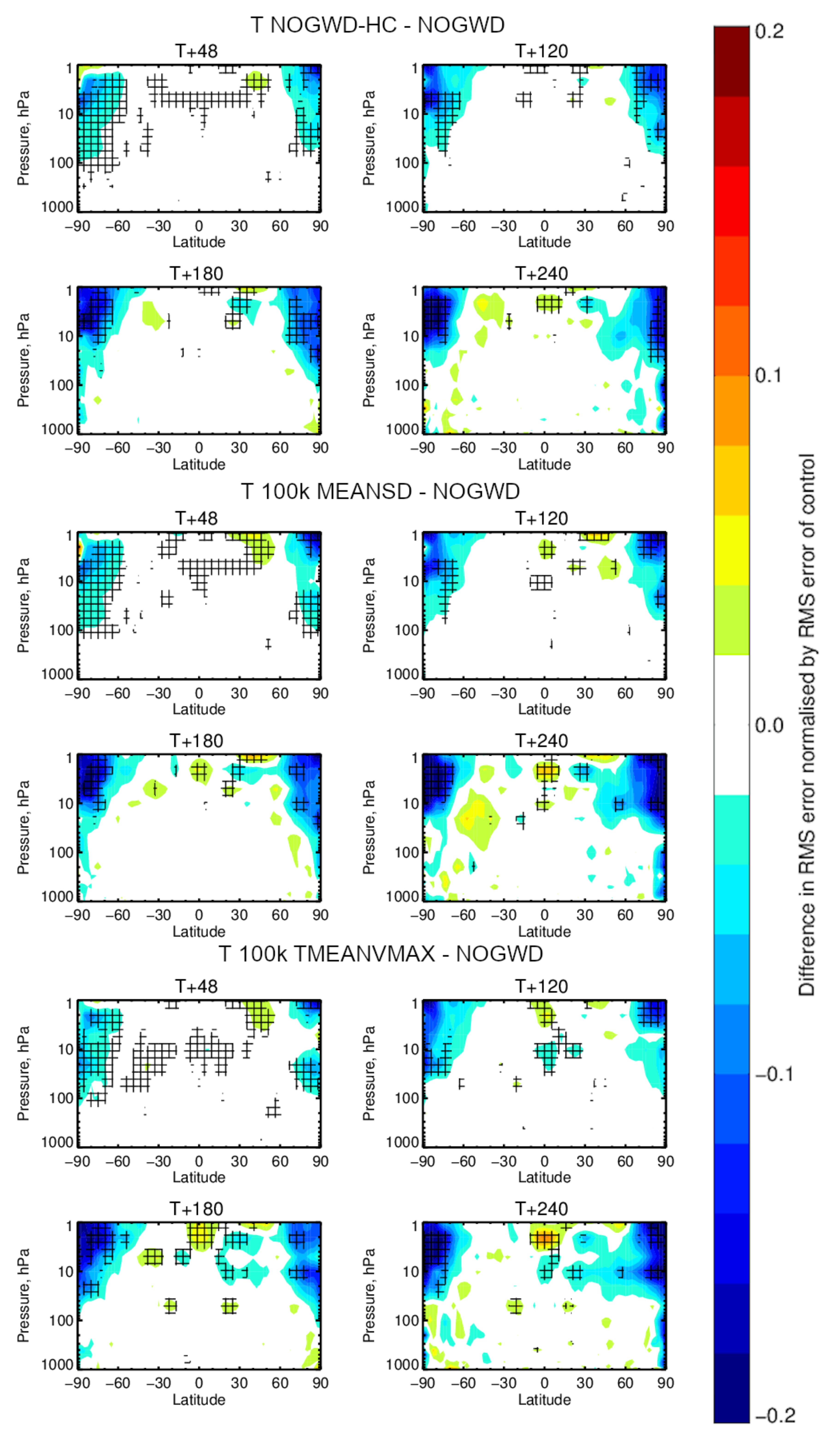}
    \caption{Change in the zonally-averaged RMS error against the analysed state of the atmosphere relative to the existing NOGWD scheme for the temperature field. Top: the change when using the high-complexity (HC) variant of the existing scheme after 48, 120, 180 \& 240 hours. Middle: equivalent changes when using our 100k MEANSD NN. Bottom: equivalent changes when using our 100k TMEANVMAX NN. Results for each are averaged over 62 forecasts corresponding to initialising the model at midnight every day in July and December 2019. Hatching denotes statistically significant differences. Significant improvements are seen for HC and NN models over the existing scheme. In particular the strong similarities between the patterns for the HC scheme and neural networks, indicating the neural network has accurately emulated the HC scheme.
    }
\label{fig:IVER_T}
\end{figure}

\begin{figure}
    \centering
    \includegraphics[width=0.7\columnwidth]{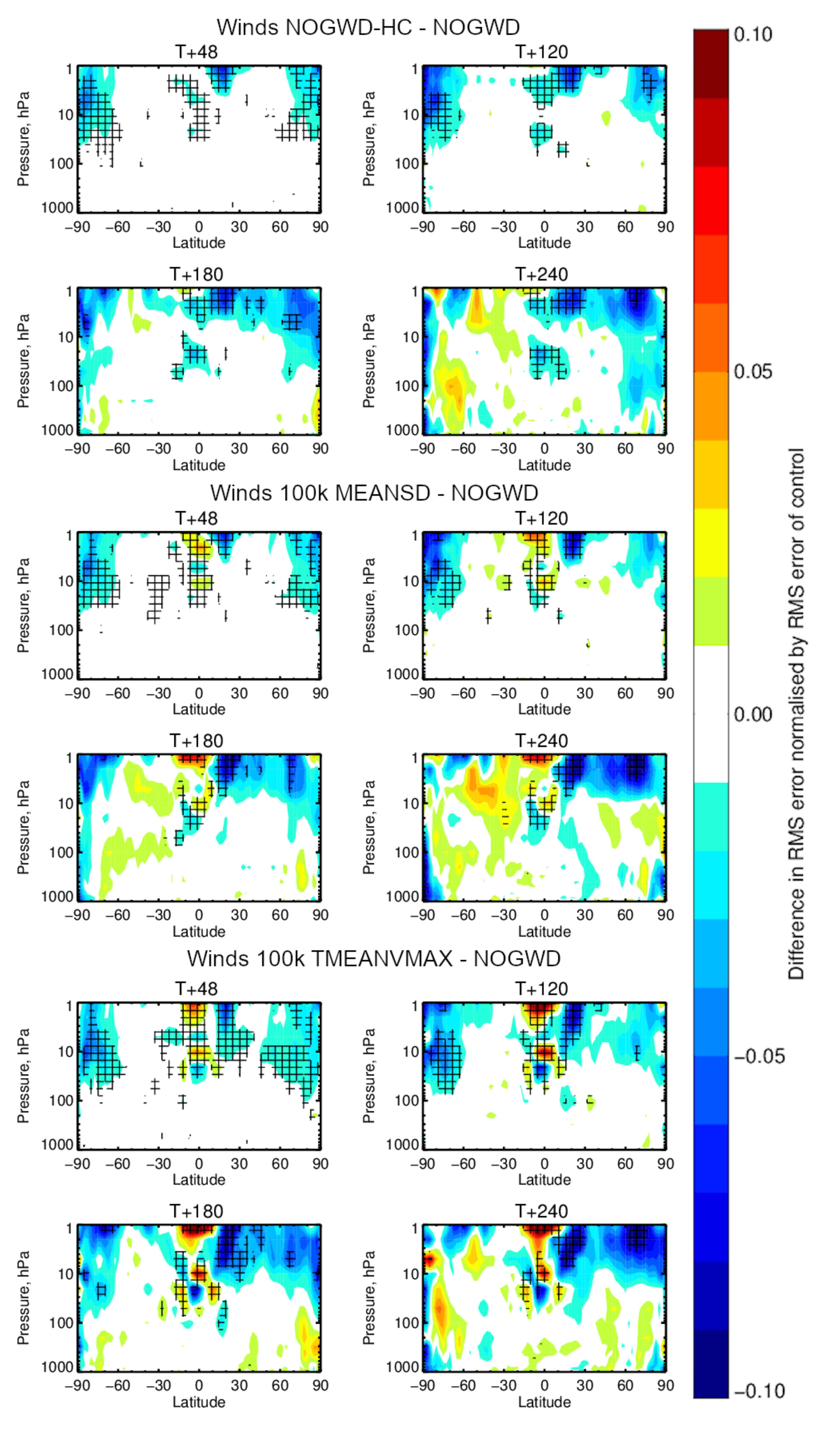}
    \caption{Equivalent to figure \ref{fig:IVER_T} but for the horizontal wind components. While more red (reduced accuracy) is observed than for the temperature field both the HC and neural network solutions show more improvements and crucially show similar patterns, indicating the neural network has accurately emulated the HC scheme. Both NN solutions introduce some degradation in the tropics around 1hPa.
    }
    \label{fig:IVER_W}
\end{figure}

{We now carry out 10-day forecasts at the TCo399 resolution starting each day within July and December 2019. In figures \ref{fig:IVER_T} and \ref{fig:IVER_W} we plot the results averaged across the start dates. We measure the difference in RMSE for pairs of forecast models, each assessed against the ECMWF operational analysis for the temperature (figure \ref{fig:IVER_T}) and wind fields (figure \ref{fig:IVER_W}). The top four panels of each plot show the improvement/degradation (blue/red contours) in forecast quality when using the HC variant of the existing NOGWD scheme versus the existing scheme. Hatching denotes statistically significant differences \cite{geer2016significance}. The other eight panels show equivalent results when using the 100k MEANSD or 100k TMEANVMAX neural networks.
For both fields we see the very similar patterns of change in forecast quality for both the HC and NN schemes, demonstrating the accuracy to which this HC scheme has been learnt.
Especially for the temperature field both the HC and NN schemes provide notable improvements in forecast quality, particularly over the poles where the error is reduced by more than 0.2C for long lead-times. For the winds, there is a balance between improvement with some degradation in the tropics around 1hPa. Overall the neural network has a positive impact on the forecast quality. 
In these tests there is no clear winner between the MEANSD and TMEANVMAX normalisation approaches. 
}

\subsection*{Year-long simulations}

\begin{figure}
    \centering
    \includegraphics[width=\columnwidth]{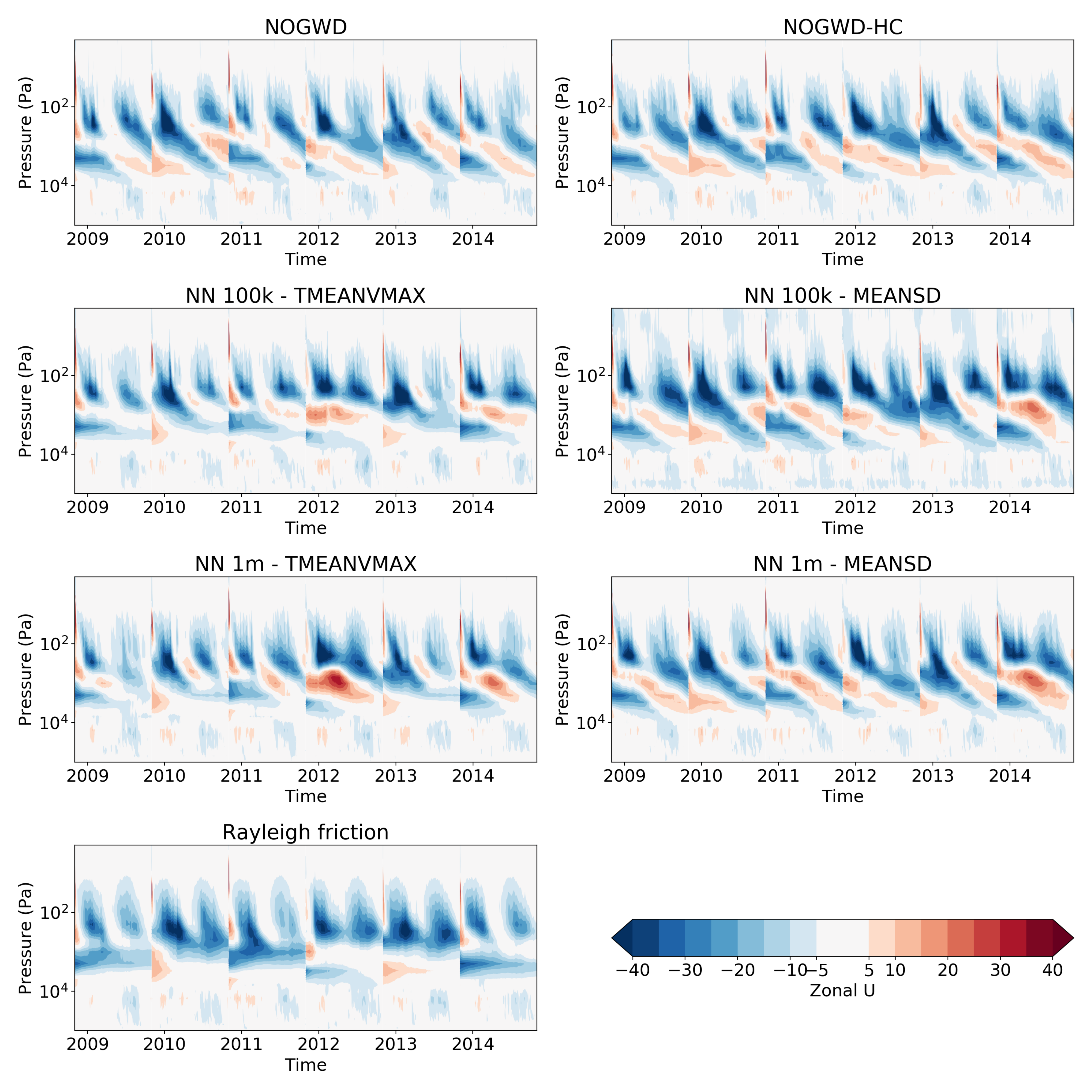}
    \caption{Average zonal-mean zonal jet between latitudes -5 to 5 for six consecutive year-long integrations at TL159 resolution (${\sim}125 \mathrm{km}$). Plots show the ability of the IFS model to capture the QBO when coupled to different NOGWD schemes: NOGWD, the current scheme; NOGWD HC, the increased complexity variant; 4 NN schemes covering different complexities and normalisation approaches; Rayleigh friction, the precursor scheme to the current NOGWD scheme. Both variations of the current scheme produce similar dynamics, which are faithfully reproduced by the MEANSD neural networks. The TMEANVAX networks outperform Rayleigh friction but fail to adequately reproduce the descent phase of the QBO.}
    \label{fig:qbo}
\end{figure}

\begin{figure}
    \centering
    \includegraphics[width=\columnwidth]{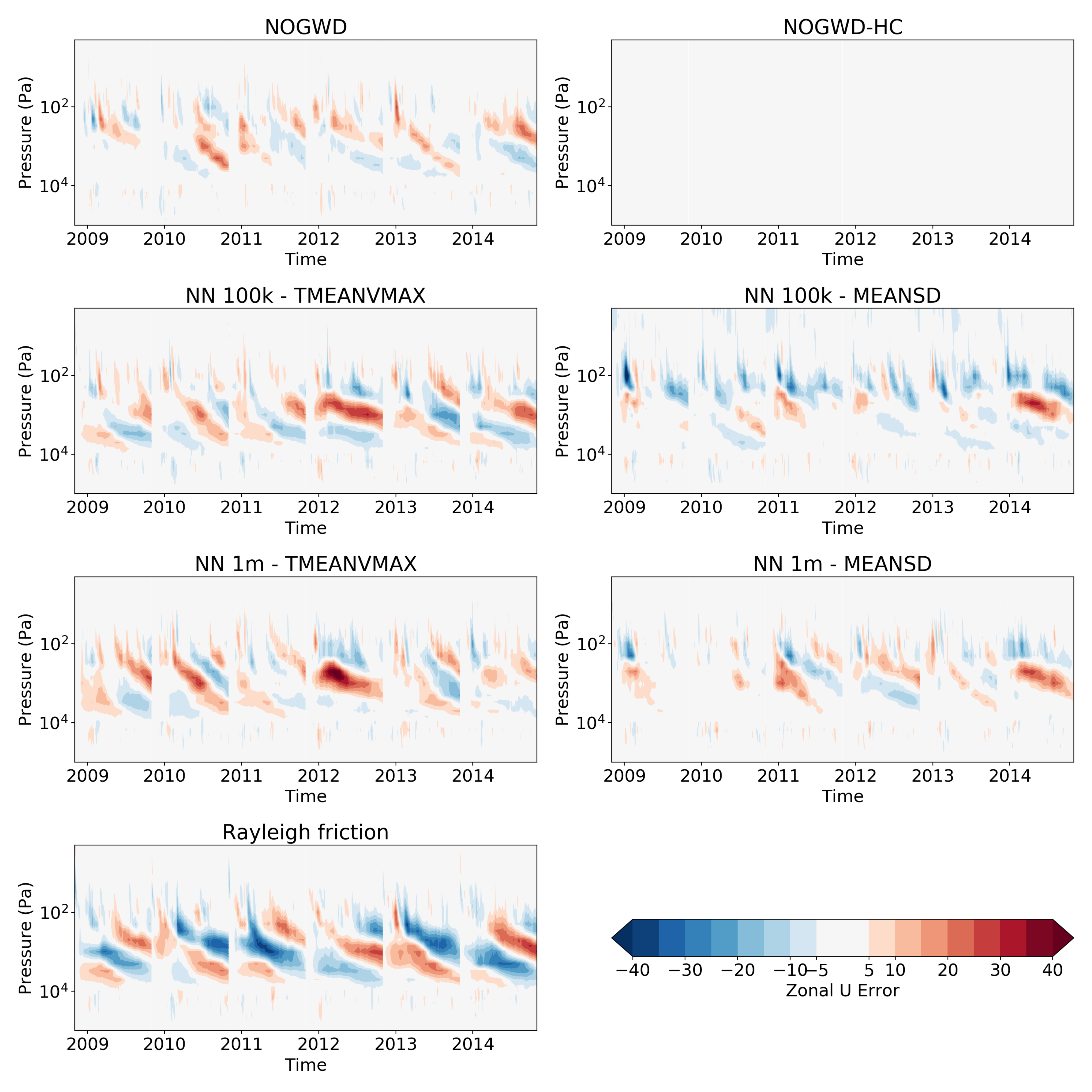}
    \caption{Difference in the zonal jets between the simulations plotted in figure \ref{fig:qbo}. NOGWD HC is treated as the reference scheme and hence has no error. Both 100,000 (100k) and 1 million (1m) DOF MEANSD schemes have comparable errors to the existing NOGWD scheme. Errors for the TMEANVMAX scheme are larger.}
    \label{fig:qboerror}
\end{figure}

\begin{figure}
    \centering
    \includegraphics[width=\columnwidth]{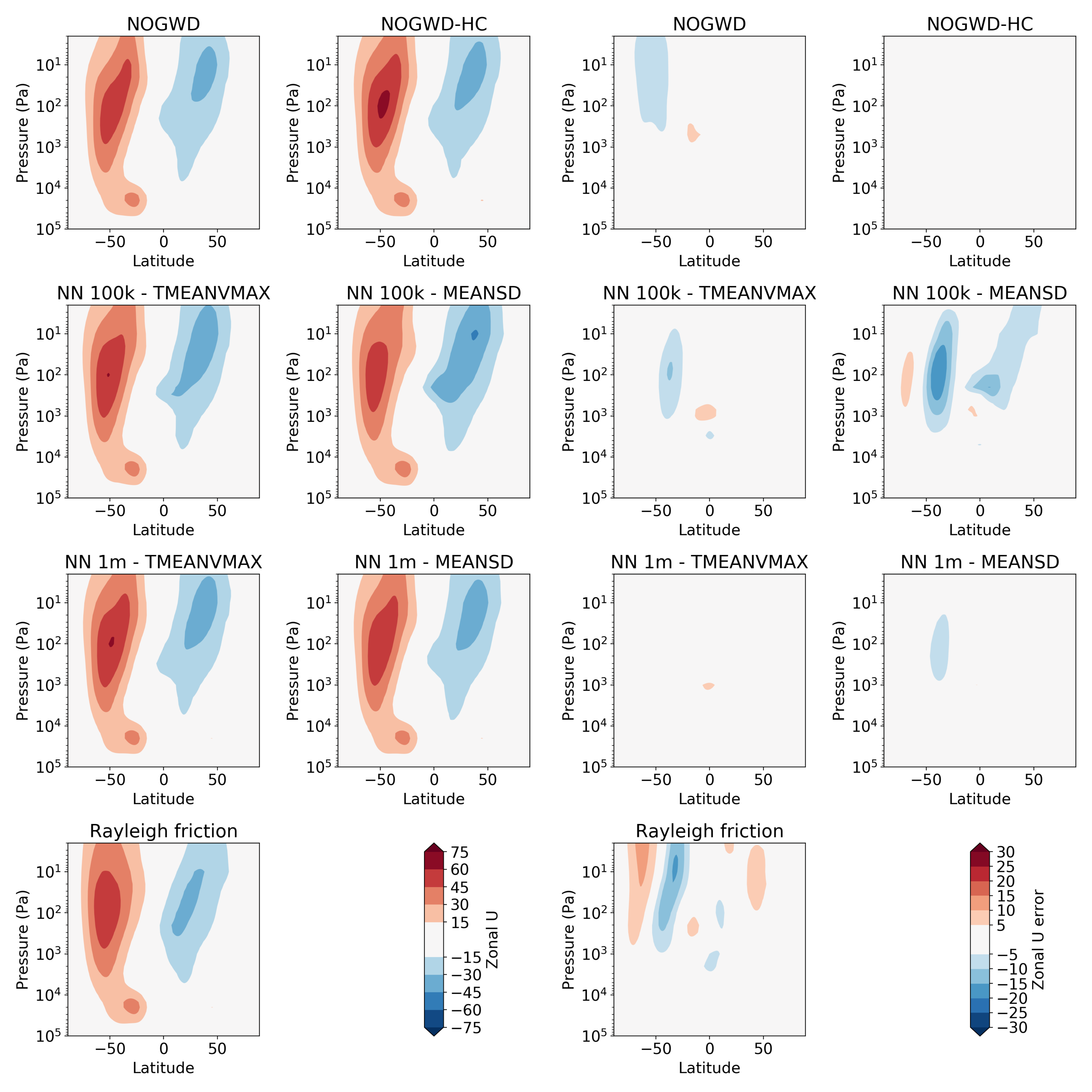}
    \caption{June-July-August (JJA) average zonal jet as a function of latitude and pressure level for variants of the NOGWD scheme. Plots on the left show the jet, with differences to the NOGWD HC scheme depicted on the right. Here the more expensive schemes slightly outperform their cheaper counterparts.}
    \label{fig:summerjet}
\end{figure}

\begin{figure}
    \centering
    \includegraphics[width=\columnwidth]{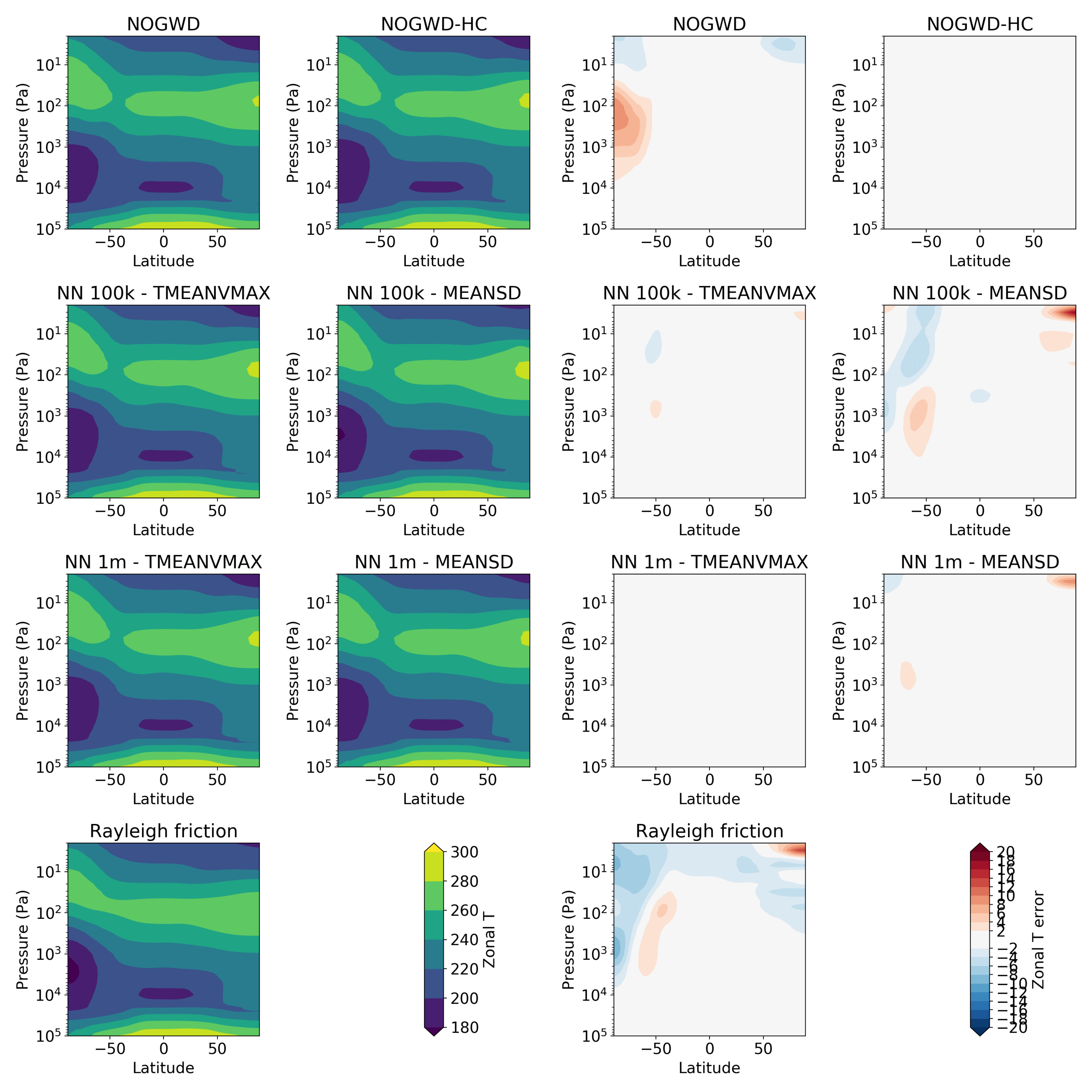}
    \caption{JJA average temperature profiles. Left plots show the temperature profiles and right plots show the deviations from the NOGWD HC result. Each scheme produces very similar temperature profile to our reference simulation. The MEANSD models have a small warm bias over the Arctic.}
    \label{fig:summerT}
\end{figure}

In order to fully measure the accuracy and stability of our neural networks we now assess their performance in long-range forecasting. When the NOGWD scheme was first introduced, it was found to have a positive impact on the ability of the IFS model to capture the quasi-biennial oscillation (QBO) when compared with the previous Rayleigh friction model \citeA{orr2010improved}. Crucially, the Rayleigh friction model fails to capture the descent phase and produces an overly-regular oscillation period.
Additionally, the latitude-pressure distribution of zonal winds and temperatures in the middle-atmosphere were improved by the introduction of the NOGWD scheme \citeA{orr2010improved}.
For computational cost reasons we carry out these simulations with TL159 (${\sim}125 \mathrm{km}$ horizontal resolution), but do not retrain for this different resolution. To examine the behaviour of our networks on long timescales we carry out six year-long simulations starting on the 1st of November in the years 2009 to 2014. We forecast this period with the high-complexity and normal versions of the existing NOGWD scheme, the Rayleigh friction scheme and four of the configurations described in the previous section, omitting the cheapest networks. In figure \ref{fig:qbo} we plot the time-pressure behaviour of the equatorial zonal wind, calculated as the horizontal average between latitudes -5 and 5.  Discontinuities indicate where a new simulation is started and exist for each experiment. Figure \ref{fig:qboerror} shows differences in jet structure from the HC truth run. Each of the neural networks tested significantly outperforms the old Rayleigh friction scheme and shows clear descent of the jet altitude. We find that the MEANSD normalised networks do a better job at capturing the strong descents observed in the two variants of the original schemes in 2013 and 2015. Interestingly, we find only a small improvement for the 1 million (1m) DOF models over the 100,000 (100k) DOF models. For the 100k MEANSD normalisation we observe a small amount of noise at the top of the atmosphere, but this appears to have no noticeable impact on the jet structure.

From the same experiments in figures \ref{fig:summerjet} and \ref{fig:summerT} we plot the June-July-August (JJA) zonal-mean zonal wind and temperature distribution, averaged over all six start-dates. This period was chosen to capture winter in the southern-hemisphere, comparable results were seen for the December-January-February period for the northern-hemisphere winter period (not plotted). For the zonal winds we see some improvement in the jet structure for the more expensive neural networks. For the temperature field we see a slightly better performance for the networks normalised using the TMEANVMAX scheme, in contrast to the results for the QBO where the MEANSD schemes produce better jet structure. 

In these yearly integrations we find some evidence that neural networks trained with the MEANSD normalisation approach outperform the TMEANVMAX method. Reexamining figure \ref{fig:tendencies} we see a local maximum in tendency activity around the 35th model level (${\sim}70\mathrm{hPa})$. In the MEANSD approach, each level has equal contribution to the loss function, encouraging our networks to capture these variations. In the TMEANVMAX approach these mid-level tendencies are an order of magnitude smaller than those at the top of the atmosphere and so have little contribution to our loss function. As such the network struggles to reproduce the descent as faithfully. 

\section{Discussion}
\label{sec:Discussion}

\subsection*{Conservation properties}
The existing NOGWD scheme conserves momentum. A fixed amount of momentum flux is launched upwards within a column, with momentum deposited at each layer depending on wave stability. All remaining momentum that reaches the upper-most layer is deposited there to ensure conservation. In our first iterations of neural networks we utilised the same approach and trained our networks to predict all other output layers and used momentum conservation to deduce the upper-most tendency. This approach is similar to the work of \citeA{beucler2019enforcing}. In our formulation, our training loss does not explicitly include the upper-most layer, whereas \citeA{beucler2019enforcing} use conservation properties to deduce the upper-most layer which is then included in the loss calculation. 
However, during our long-range forecasting experiments we found that this momentum conservation approach occasionally destabilised our simulations. 
Due to the very strong difference in mass per volume of air, a correction of momentum at the top of the model, that is caused by a relatively small change in momentum close to the ground, can have a significant detrimental effect on the local momentum budget at the model top which can trigger model instabilities.
For our final networks we abandon exact momentum conservation and instead directly predict the top layer tendencies along with all other layers using our networks. With this approach we are able to produce stable and accurate parameterisation schemes. Further work could be undertaken to follow  \citeA{beucler2019enforcing} and use conservation of momentum before the loss is calculated to test if this stabilises the forecasts.

\subsection*{Performance analysis}

Given our motivation for building these networks i.e. improving the overall performance of the IFS model, an assessment of the performance is a crucial albeit complex step. As previously highlighted, the current NOGWD scheme uses approximately $100,000$ FLOPs to calculate the NOGWD tendencies for a single column. We test our performance in 10 day forecasts at TCo399 resolution, using 66 processors each using 6 threads. The current NOGWD scheme is run at the IFS standard of double-precision, whereas the NNs are run at the trained precision, single-precision. In this setup 1.6 million columns are evaluated per thread during the simulation. In our test IFS setup 300 seconds or $1\%$ of the total simulation time are spent on the NOGWD calculations. By comparison, when using one of our 100k DOF models in the equivalent setup the NOGWD scheme takes 360 seconds per thread, slightly slower than the current version. For a 30k DOF model, this is reduced to 250 seconds, slightly cheaper than the current scheme but an insignificant performance increase. To better understand the performance bottlenecks of both schemes we also run each scheme decoupled from the IFS model. Simulating the same number of columns (1.6 million) in decoupled testing takes 18 seconds for the existing NOGWD scheme, 20 seconds for a 100k DOF model or 10 seconds for a 30k DOF model. This shows that both the original and neural network scheme are heavily limited by data movement, so approaches that can overcome this problem can have dramatic performance increases. One such vision is a heterogeneous hardware cluster where some of the parameterisation schemes are calculated on GPUs with the output being sent back to the CPU for the remaining simulations. We have shown here that this could be achieved by training neural network emulators, meaning that existing code does not have to be converted to CUDA or other GPU-appropriate languages. To understand the possible benefits of this architecture we test our neural networks on a NVidia V100 GPU. Even when including the time for transferring the data to and from the GPU a 100k DOF NN takes 4 seconds to simulate 1.6million columns. Even this time is dominated by overhead costs, as a 1m DOF NN also takes 4 seconds. Therefore, while the CPU performance when coupled to the IFS is not currently faster than the existing scheme, there is scope for dramatic performance gains on heterogeneous hardware, particularly if more parameterisation schemes can be accurately emulated by neural networks. {ECMWF's scalability program \cite{bauer2020ecmwf} is currently adapting existing parametrisation schemes to be GPU portable, enabling a GPU comparison in the future.}

\subsection*{Reduced numerical precision}

Recent developments of GPU and Tensor Processing Unit (TPU) hardware have given users access to low numerical precision floating point numbers with significantly improved performance for neural networks. 
Outside of machine learning, elements of both the dynamics \cite{hatfield2019accelerating} and parameterised physics \cite{saffin2020reduced} can be calculated at half-precision with no impact on forecast quality.
To test the applicability of reduced precision networks for our emulation we utilise Tensorflow's mixed-precision capability which stores variables at 32 bits (single-precision) but uses 16 bits (half-precision) for intermediate calculations. Through both our training and offline testing phases we find no notable degradation in the accuracy of our networks. 
{Currently, most CPU architecture does not support calculations at half-precision,}
so online testing cannot currently be easily carried out. 
Recent work in emulation of convection parameterisation used emulated half-precision with good network accuracy \cite{yuval2020use}. It is future work to couple our emulators to the IFS model in such a way that GPUs and reduced numerical precision can be leveraged. 

\subsection*{Alternate network structures}

\begin{figure}
    \centering
    \includegraphics[width=\columnwidth]{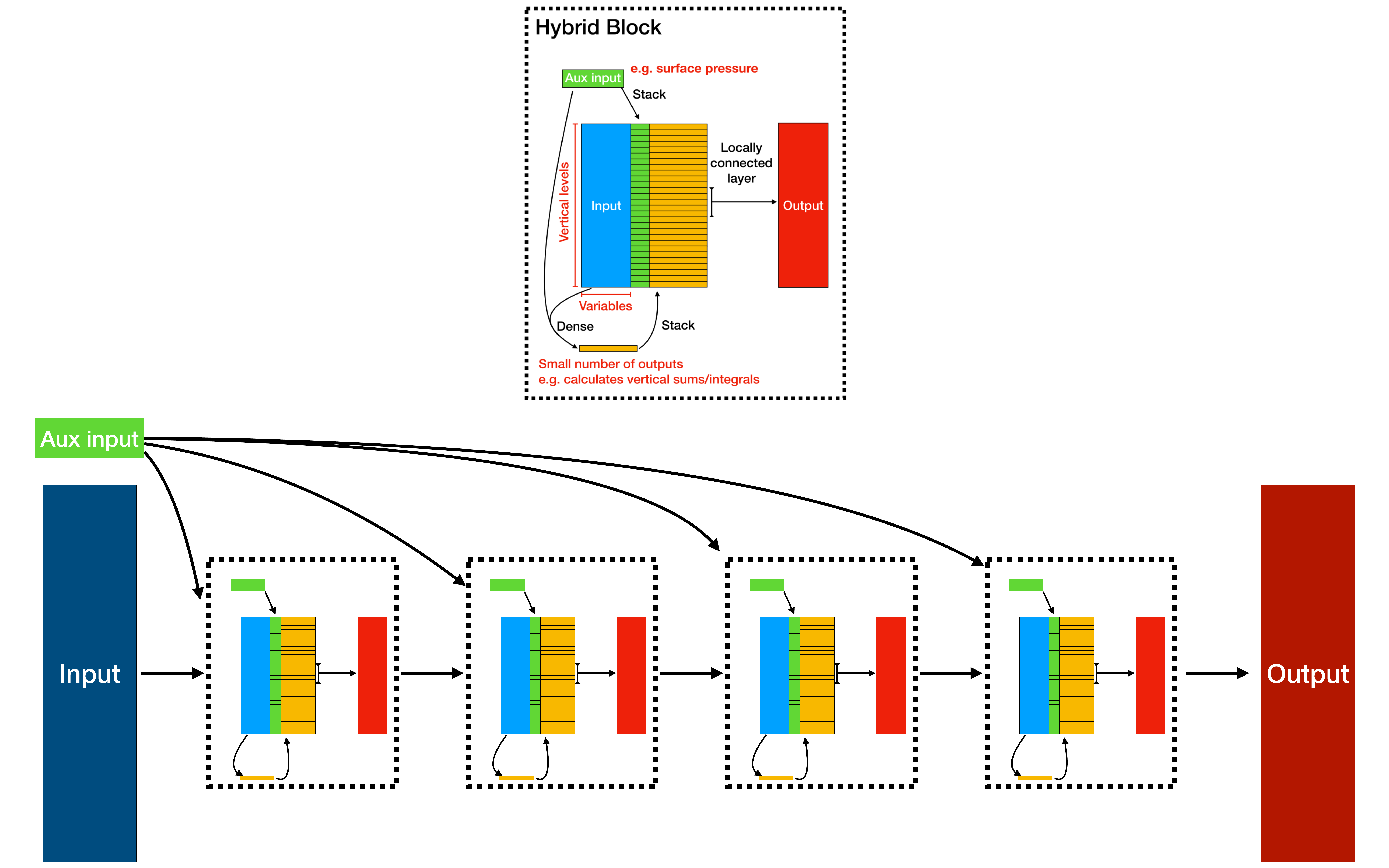}\\
    \caption{Schematic of our hybrid convolution deep network. Top: our hybrid block. Bottom: An Example model layout consisting of four blocks. The Input block to the hybrid networks are the profiles of winds and temperature. The Aux input refers to the surface pressure and geopotential, which are injected to each vertical layer of each block. Arrows indicate layers, e.g. a dense (fully-connected) layer taking the inputs and aux inputs. Stack indicates a repeating of an object to enable concatenation with data that has vertical structure (e.g. wind profiles.)
    }
    \label{fig:schematic}
\end{figure}

In this study we considered fully-connected fixed-width neural networks, one of the simplest network designs available. In this architecture each node in the first layer has access to all input features, with no knowledge of the vertical structure of the atmosphere. Given that many calculations in physics are local in space, e.g. calculation of a vertical derivative, this suggests that improved performance could be found by a more appropriate choice of network architecture. 

One choice is using convolutional layers, a fixed local (in the vertical direction) operator where the same weights are used for each vertical layer in the atmosphere. Given that our model levels are unequally spaced in both distance from the Earth (from O(m) to O(km) distance per layer) and pressure values this appears to be a poor choice for our application. When testing with fully convolutional networks we find that offline errors are dramatically higher than for our fully connected networks, irrespective of the number of filters or convolutional layers used. 

An alternate choice is to use locally-connected layers. These have the same stencil shape as convolution layers, but the weights are trained individually for each vertical layer. It is therefore possible to encode a finite difference vertical derivative on our vertical grid within this architecture. Despite this, our testing finds that models comprising of entirely locally-connected layers still fail to match the performance of our fully-connected networks, with testing errors more than 2 times larger than our best 100k DOF fully-connected networks.

Examining the algorithm of the existing NOGWD scheme we can understand why purely local approaches struggle to achieve good predictive performance. In the existing scheme, while most of the calculations are local in vertical space, there are several points where non-local calculations take place, for example when velocities relative to that of the launch level wind velocities are calculated. With a purely local method, information can only propagate through the network based upon the filter width at each layer, so with a width of 5, a minimum of 12 layers would be required for information to be able to propagate from the lowest to highest vertical layer.

\begin{figure}
    \centering
    \includegraphics[width=0.55\columnwidth]{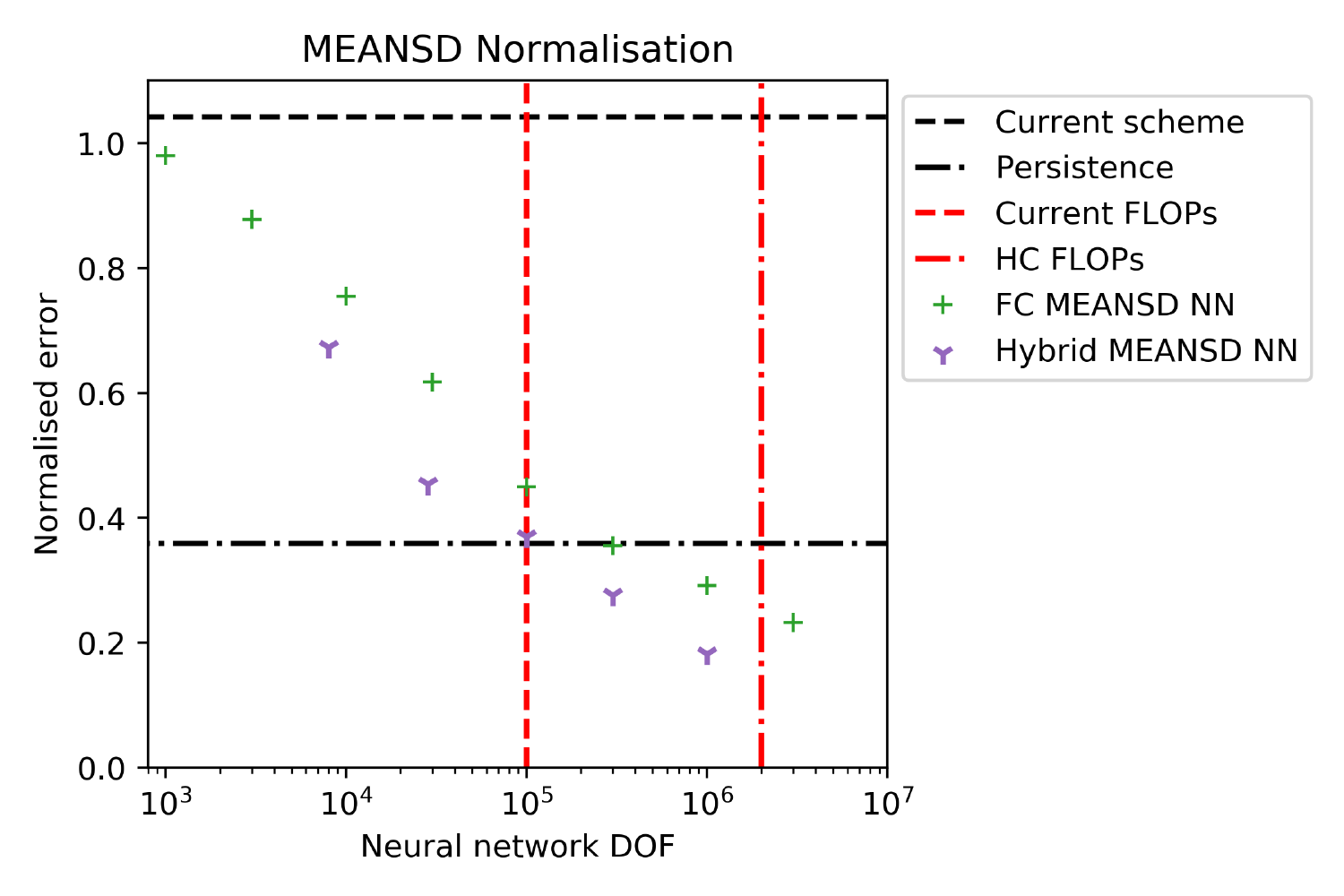}\\
    \caption{Offline results on the test dataset using the MEANSD normalisation method. In addition to the results presented in figure \ref{fig:offline} we plot our best performing hybrid networks as outlined in figure \ref{fig:schematic}. Consistently the hybrid network outperforms the fully-connected networks, with approximately three-times reduction in the required DOF for an equivalent test error.}
    
    \label{fig:offline2}
\end{figure}

With the above in mind we design a hybrid approach, utilising both fully-connected and locally-connected layers, a schematic of which is plotted in figure \ref{fig:schematic}. Each block of this network comprises of a small number of dense connections, which can combine both the inputs to that layer and the auxiliary inputs, which in this context are surface pressure and surface geopotential.
The output of these dense layers are then stacked vertically, alongside auxiliary inputs. All of these features are then passed through a set of locally-connected filters. This design seeks to predominantly leverage the local nature of the physical parameterisation scheme while also enabling information to propagate throughout the domain within a single block. In figure \ref{fig:offline2} we re-plot the results for the fully-connected networks using MEANSD normalisation and add the results of our hybrid networks. 
To explore the hyper-parameters associated with our hybrid networks (e.g. the number of blocks or number of dense outputs) we use the HyperOpt tool \cite{bergstra2013making}.
For all DOF values our hybrid networks significantly outperform their fully-connected equivalents, producing comparable results to a fully-connected network with three-times the DOF. There is certainly room for further improvements in searching this and other network architecture spaces. Due to the increased network architecture complexity we have not yet implemented these networks within our Fortran approach. This network architecture could prove to be a useful approach across the spectrum of physical parameterisation schemes.

\subsection*{Generalisability to model resolution and version changes}

\begin{figure}
    \centering
    \includegraphics[width=\columnwidth]{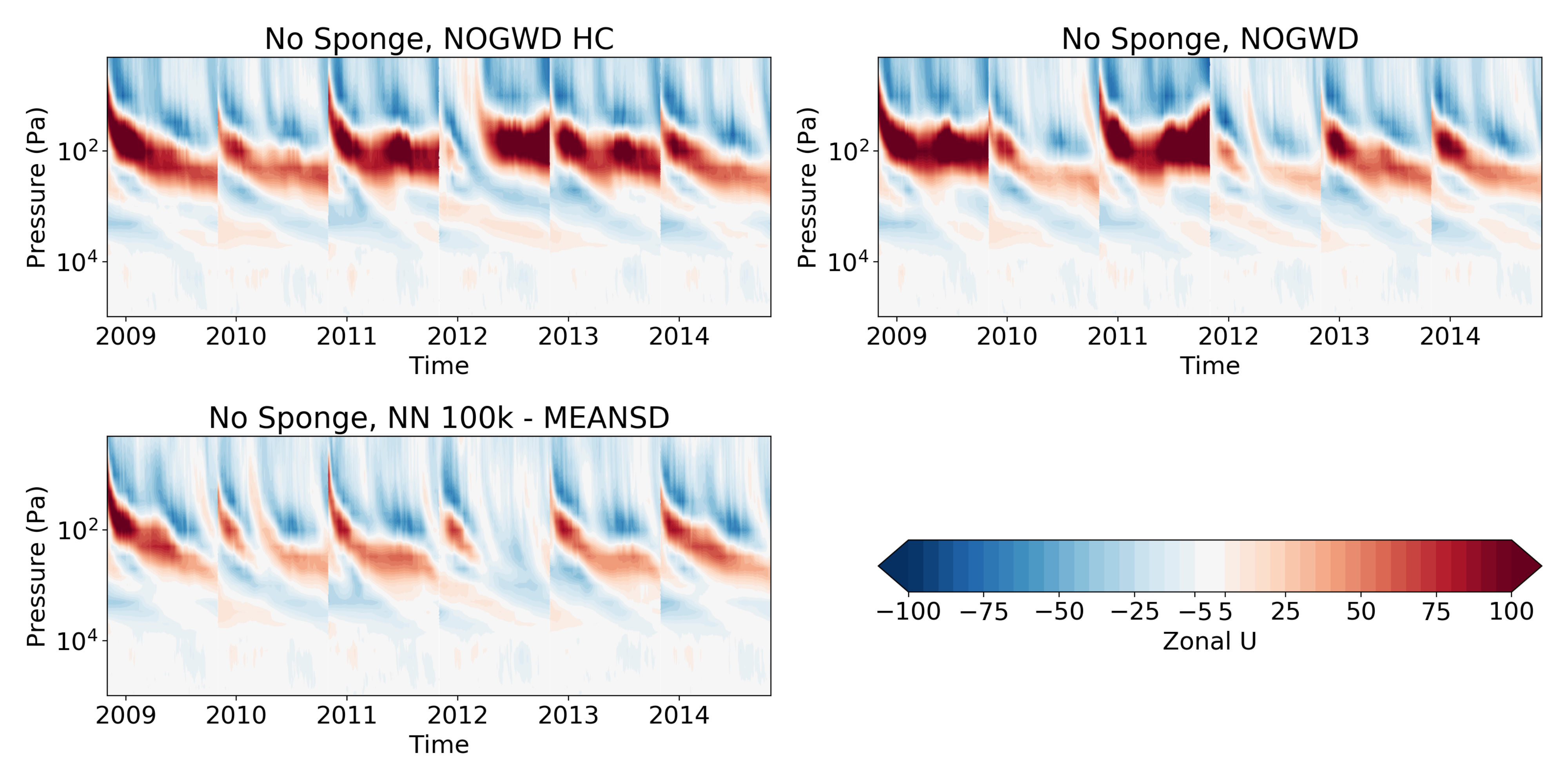}\\
    \caption{Average zonal-mean zonal jet between latitudes -5 to 5 for six consecutive year-long integrations at TL159 resolution (${\sim}125 \mathrm{km}$). Experiments match equivalent names from figure \ref{fig:qbo} except that the numerical sponge at the top of the atmosphere has been removed. Wind speeds in the stratosphere are significantly increased, yet the QBO phase and amplitude is captured by the NN.
    }
    
    \label{fig:sponge}
\end{figure}

If a body of work is undertaken to produce emulators of physical parameterisaton schemes within an operational modelling system it is important to understand possible sensitivities to changes in either the model resolution or model version. If any changes of these hyperparameters necessitated the learning of new neural networks, and hence generation of new training datasets, then this could quickly become a very large infrastructure project to maintain. Vertical resolution is an intrinsic part of each parameterisation scheme, and so changes to vertical resolution will require generation of new schemes. However for horizontal resolution, in the context of the NOGWD scheme, we found no sensitivity to changing the spatial resolution 
{between 25km and 125km.  Our NNs are able to generalise across the subtle changes in atmospheric profiles seen across these resolutions}. We tested this both offline and online, e.g. our results for long-range forecasting are simulated at a lower spatial resolution than the training dataset. For model cycle upgrades, we tested performance across three consecutive cycles of IFS (45r1,46r1 \& 47r1) and found no noticeable degradation in performance. In figure \ref{fig:sponge} we show the evolution of the equatorial zonal-mean zonal jet with the sponge layer removed. Despite no training on a dataset with the sponge removed our MEANSD NN produces a stable atmospheric structure. This is despite a significant increase in stratospheric wind speeds (note the change in contour levels compared to figure \ref{fig:qbo}. The phase and amplitude of the QBO is comparable to our simulation with the HC NOGWD scheme with no sponge layer.
This is more evidence that our neural networks show reasonable robustness to model changes. However, we cannot rule out degradation in scenarios where very significant model changes are made.

\subsection*{Orographic gravity wave drag}

Armed with the successes of our emulation of NOGWD we carried out an equivalent task to learn the orographic gravity wave drag (OGWD) scheme within IFS. We used the same methodology for data generation, same normalisation methods and same network architecture choices. Orographic gravity wave drag shares the same inputs as its non-orographic counterpart, with the addition of four parameters describing the unresolved orography. The existing OGWD scheme is active only in locations with significant unresolved orography and when the atmospheric profile is susceptible to the waves generated by that orography. The result of this is a scheme that is inactive for most of the globe and is often inactive even in locations with significant unresolved orography. 

Unfortunately, our attempts to learn this OGWD scheme with fully-connected neural networks were unsuccessful. Increasing the degrees of freedom did not reduce the testing error, unlike our results for the NOGWD scheme.  
Our neural networks struggled to derive this nuanced, yet computationally cheap (${\sim}20,000$ FLOPs per column), interaction between the orography and atmospheric profile. 
The most common result of training was networks that predicted zero velocity tendency for almost all input columns. In online testing our model performed similarly well to a forecast without an OGWD scheme. 
In attempting to overcome these problems we tested many approaches. Learning a single grid-point, i.e. generating a network which could predict the tendencies for only a single point in space, was possible but is not a scalable solution. 
The existing OGWD scheme is only utilised when sub-grid orographic variability exceeds a threshold, we followed this example and trained only on locations where this threshold was met.
Additionally we tested methods to change the balance of training data to exclude a proportion of our training data where the scheme was inactive, but this often resulted in a dramatic increase in false positives (OGWD activity where the current scheme is inactive).

From this experience we summarise that training emulators of physical parameterisation schemes is a non-trivial task, not guaranteed to succeed even with large amounts of available data. Our outlook is that most parameterisation schemes share more in common with the NOGWD scheme and therefore will be possible to learn. However this remains to be tested. 

\section{Conclusions}

In this work we successfully emulate the non-orographic gravity wave drag scheme from the operational IFS forecasting model. Despite training offline we are able to produce an emulator which can run stably when coupled to the IFS for seasonal timescales. The most exacting test of our emulators, reproducing the descent of the QBO, was successfully achieved by our neural networks. {We have tested two different normalisation approaches that, interestingly, showed similar performance across our tests.}
Broadly, our networks have similar computational cost to the existing scheme when coupled to the IFS on a CPU-based architecture. However, both the existing scheme and our neural networks are heavily limited by data movement when run within the IFS. The major advantage of our networks is the ease of portability to heterogeneous architectures, where neural network emulators of parameterisation schemes could be offloaded to GPUs.
This approach will have the greatest benefit if many parameterisation schemes can be emulated with neural networks. However, as discussed above for the orographic gravity wave drag, other schemes might prove more challenging to emulate as accurately as the NOGWD scheme. 
Beyond forecasting, there are other values of building parameterisation emulation. The 4D-var data assimilation approach uses a tangent-linear and adjoint to the forecasting model, constructed of tangent-linear and adjoint versions of each kernel. This is a challenging process typically derived by hand. However, with an accurate neural network emulator the tangent linear and adjoint versions can be trivially derived due to the simple nature of a neural network's algorithm. In our sister paper \citeA{hatfieldTL} we will test how these tangent linear and adjoint versions perform within a data-assimilation framework.
{Further exploration of the performance of these emulators would be required before one could be considered for operational forecasting. This would include high-resolution testing (9km), seasonal forecast tests, ensemble forecasts and using the emulator within the nonlinear integrations within 4D-var.}

\acknowledgments
The data used to build the machine learning emulators has been archived and is freely accessible at \url{https://doi.org/10.5281/zenodo.4740758} .

M. Chantry and T. Palmer were supported by grant DCR00481 from the Office of Naval Research Global. ECMWF provided computing resources resources, including the European Weather Cloud, which is an ECMWF and EUMETSAT project. Thanks to Ioan Hadade for his technical support. P. Dueben gratefully acknowledges funding from the Royal Society for his University Research Fellowship as well as the AI4Copernicus and MAELSTROM projects that are funded under Horizon 2020 and the European High-Performance Computing Joint Undertaking (JU; grant agreement No 101016798 and 955513). The JU receives support from the European Union’s Horizon 2020 research and innovation programme and United Kingdom, Germany, Italy, Luxembourg, Switzerland, Norway. S. Hatfield and P. Dueben have also received funding from the ESIWACE Horizon 2020 project (grant agreement No 823988). T. Palmer also received funding from an ERC Advanced Grant, ITHACA 741112.


\begin{thebibliography}{}

\bibitem [\protect \citeauthoryear {%
Abadi%
\ \protect \BOthers {.}}{%
Abadi%
\ \protect \BOthers {.}}{%
{\protect \APACyear {2015}}%
}]{%
tensorflow}
\APACinsertmetastar {%
tensorflow}%
\begin{APACrefauthors}%
Abadi, M.%
, Agarwal, A.%
, Barham, P.%
, Brevdo, E.%
, Chen, Z.%
, Citro, C.%
\BDBL {}Zheng, X.%
\end{APACrefauthors}%
\unskip\
\newblock
\APACrefYearMonthDay{2015}{}{}.
\newblock
\APACrefbtitle {{TensorFlow}: Large-Scale Machine Learning on Heterogeneous
  Systems.} {{TensorFlow}: Large-scale machine learning on heterogeneous
  systems.}
\newblock
\begin{APACrefURL} \url{https://www.tensorflow.org/} \end{APACrefURL}
\newblock
\APACrefnote{Software available from tensorflow.org}
\PrintBackRefs{\CurrentBib}

\bibitem [\protect \citeauthoryear {%
Bauer%
\ \protect \BOthers {.}}{%
Bauer%
\ \protect \BOthers {.}}{%
{\protect \APACyear {2020}}%
}]{%
bauer2020ecmwf}
\APACinsertmetastar {%
bauer2020ecmwf}%
\begin{APACrefauthors}%
Bauer, P.%
, Quintino, T.%
, Wedi, N.%
, Bonanni, A.%
, Chrust, M.%
, Deconinck, W.%
\BDBL {}others%
\end{APACrefauthors}%
\unskip\
\newblock
\APACrefYear{2020}.
\newblock
\APACrefbtitle {The ECMWF Scalability Programme: Progress and Plans} {The ecmwf
  scalability programme: Progress and plans}.
\newblock
\APACaddressPublisher{}{European Centre for Medium Range Weather Forecasts}.
\PrintBackRefs{\CurrentBib}

\bibitem [\protect \citeauthoryear {%
Bergstra%
, Yamins%
\BCBL {}\ \BBA {} Cox%
}{%
Bergstra%
\ \protect \BOthers {.}}{%
{\protect \APACyear {2013}}%
}]{%
bergstra2013making}
\APACinsertmetastar {%
bergstra2013making}%
\begin{APACrefauthors}%
Bergstra, J.%
, Yamins, D.%
\BCBL {}\ \BBA {} Cox, D.%
\end{APACrefauthors}%
\unskip\
\newblock
\APACrefYearMonthDay{2013}{}{}.
\newblock
{\BBOQ}\APACrefatitle {Making a science of model search: Hyperparameter
  optimization in hundreds of dimensions for vision architectures} {Making a
  science of model search: Hyperparameter optimization in hundreds of
  dimensions for vision architectures}.{\BBCQ}
\newblock
\BIn{} \APACrefbtitle {International conference on machine learning}
  {International conference on machine learning}\ (\BPGS\ 115--123).
\PrintBackRefs{\CurrentBib}

\bibitem [\protect \citeauthoryear {%
Beucler%
\ \protect \BOthers {.}}{%
Beucler%
\ \protect \BOthers {.}}{%
{\protect \APACyear {2019}}%
}]{%
beucler2019enforcing}
\APACinsertmetastar {%
beucler2019enforcing}%
\begin{APACrefauthors}%
Beucler, T.%
, Pritchard, M.%
, Rasp, S.%
, Ott, J.%
, Baldi, P.%
\BCBL {}\ \BBA {} Gentine, P.%
\end{APACrefauthors}%
\unskip\
\newblock
\APACrefYearMonthDay{2019}{}{}.
\newblock
{\BBOQ}\APACrefatitle {Enforcing analytic constraints in neural-networks
  emulating physical systems} {Enforcing analytic constraints in
  neural-networks emulating physical systems}.{\BBCQ}
\newblock
\APACjournalVolNumPages{arXiv preprint arXiv:1909.00912}{}{}{}.
\PrintBackRefs{\CurrentBib}

\bibitem [\protect \citeauthoryear {%
Blackford%
\ \protect \BOthers {.}}{%
Blackford%
\ \protect \BOthers {.}}{%
{\protect \APACyear {2002}}%
}]{%
blas}
\APACinsertmetastar {%
blas}%
\begin{APACrefauthors}%
Blackford, L\BPBI S.%
, Petitet, A.%
, Pozo, R.%
, Remington, K.%
, Whaley, R\BPBI C.%
, Demmel, J.%
\BDBL {}others%
\end{APACrefauthors}%
\unskip\
\newblock
\APACrefYearMonthDay{2002}{}{}.
\newblock
{\BBOQ}\APACrefatitle {An updated set of basic linear algebra subprograms
  (BLAS)} {An updated set of basic linear algebra subprograms (blas)}.{\BBCQ}
\newblock
\APACjournalVolNumPages{ACM Transactions on Mathematical
  Software}{28}{2}{135--151}.
\PrintBackRefs{\CurrentBib}

\bibitem [\protect \citeauthoryear {%
Brenowitz%
\ \BBA {} Bretherton%
}{%
Brenowitz%
\ \BBA {} Bretherton%
}{%
{\protect \APACyear {2018}}%
}]{%
brenowitz2018prognostic}
\APACinsertmetastar {%
brenowitz2018prognostic}%
\begin{APACrefauthors}%
Brenowitz, N\BPBI D.%
\BCBT {}\ \BBA {} Bretherton, C\BPBI S.%
\end{APACrefauthors}%
\unskip\
\newblock
\APACrefYearMonthDay{2018}{}{}.
\newblock
{\BBOQ}\APACrefatitle {Prognostic validation of a neural network unified
  physics parameterization} {Prognostic validation of a neural network unified
  physics parameterization}.{\BBCQ}
\newblock
\APACjournalVolNumPages{Geophysical Research Letters}{45}{12}{6289--6298}.
\PrintBackRefs{\CurrentBib}

\bibitem [\protect \citeauthoryear {%
Brenowitz%
\ \BBA {} Bretherton%
}{%
Brenowitz%
\ \BBA {} Bretherton%
}{%
{\protect \APACyear {2019}}%
}]{%
brenowitz2019spatially}
\APACinsertmetastar {%
brenowitz2019spatially}%
\begin{APACrefauthors}%
Brenowitz, N\BPBI D.%
\BCBT {}\ \BBA {} Bretherton, C\BPBI S.%
\end{APACrefauthors}%
\unskip\
\newblock
\APACrefYearMonthDay{2019}{}{}.
\newblock
{\BBOQ}\APACrefatitle {Spatially extended tests of a neural network
  parametrization trained by coarse-graining} {Spatially extended tests of a
  neural network parametrization trained by coarse-graining}.{\BBCQ}
\newblock
\APACjournalVolNumPages{Journal of Advances in Modeling Earth
  Systems}{11}{8}{2728--2744}.
\PrintBackRefs{\CurrentBib}

\bibitem [\protect \citeauthoryear {%
Brenowitz%
\ \protect \BOthers {.}}{%
Brenowitz%
\ \protect \BOthers {.}}{%
{\protect \APACyear {2020}}%
}]{%
brenowitz2020machine}
\APACinsertmetastar {%
brenowitz2020machine}%
\begin{APACrefauthors}%
Brenowitz, N\BPBI D.%
, Henn, B.%
, McGibbon, J.%
, Clark, S\BPBI K.%
, Kwa, A.%
, Perkins, W\BPBI A.%
\BDBL {}Bretherton, C\BPBI S.%
\end{APACrefauthors}%
\unskip\
\newblock
\APACrefYearMonthDay{2020}{}{}.
\newblock
{\BBOQ}\APACrefatitle {Machine Learning Climate Model Dynamics: Offline versus
  Online Performance} {Machine learning climate model dynamics: Offline versus
  online performance}.{\BBCQ}
\newblock
\APACjournalVolNumPages{arXiv preprint arXiv:2011.03081}{}{}{}.
\PrintBackRefs{\CurrentBib}

\bibitem [\protect \citeauthoryear {%
Chantry%
, Christensen%
, D{\"u}ben%
\BCBL {}\ \BBA {} Palmer%
}{%
Chantry%
\ \protect \BOthers {.}}{%
{\protect \APACyear {2021}}%
}]{%
chantryMLintro}
\APACinsertmetastar {%
chantryMLintro}%
\begin{APACrefauthors}%
Chantry, M.%
, Christensen, H.%
, D{\"u}ben, P.%
\BCBL {}\ \BBA {} Palmer, T.%
\end{APACrefauthors}%
\unskip\
\newblock
\APACrefYearMonthDay{2021}{}{}.
\newblock
{\BBOQ}\APACrefatitle {Opportunities and challenges for machine learning in
  weather and climate modelling: hard, medium and soft {AI}} {Opportunities and
  challenges for machine learning in weather and climate modelling: hard,
  medium and soft {AI}}.{\BBCQ}
\newblock
\APACjournalVolNumPages{Philosophical Transactions A}{}{}{}.
\PrintBackRefs{\CurrentBib}

\bibitem [\protect \citeauthoryear {%
Chevallier%
, Ch{\'e}ruy%
, Scott%
\BCBL {}\ \BBA {} Ch{\'e}din%
}{%
Chevallier%
\ \protect \BOthers {.}}{%
{\protect \APACyear {1998}}%
}]{%
chevallier1998neural}
\APACinsertmetastar {%
chevallier1998neural}%
\begin{APACrefauthors}%
Chevallier, F.%
, Ch{\'e}ruy, F.%
, Scott, N.%
\BCBL {}\ \BBA {} Ch{\'e}din, A.%
\end{APACrefauthors}%
\unskip\
\newblock
\APACrefYearMonthDay{1998}{}{}.
\newblock
{\BBOQ}\APACrefatitle {A neural network approach for a fast and accurate
  computation of a longwave radiative budget} {A neural network approach for a
  fast and accurate computation of a longwave radiative budget}.{\BBCQ}
\newblock
\APACjournalVolNumPages{Journal of Applied Meteorology}{37}{11}{1385--1397}.
\PrintBackRefs{\CurrentBib}

\bibitem [\protect \citeauthoryear {%
Dijkstra%
, Petersik%
, Hernandez-Garcia%
\BCBL {}\ \BBA {} Lopez%
}{%
Dijkstra%
\ \protect \BOthers {.}}{%
{\protect \APACyear {2019}}%
}]{%
dijkstra2019application}
\APACinsertmetastar {%
dijkstra2019application}%
\begin{APACrefauthors}%
Dijkstra, H.%
, Petersik, P.%
, Hernandez-Garcia, E.%
\BCBL {}\ \BBA {} Lopez, C.%
\end{APACrefauthors}%
\unskip\
\newblock
\APACrefYearMonthDay{2019}{}{}.
\newblock
{\BBOQ}\APACrefatitle {The application of Machine Learning Techniques to
  improve El Nino prediction skill} {The application of machine learning
  techniques to improve el nino prediction skill}.{\BBCQ}
\newblock
\APACjournalVolNumPages{Frontiers in Physics}{7}{}{153}.
\PrintBackRefs{\CurrentBib}

\bibitem [\protect \citeauthoryear {%
Dunkerton%
}{%
Dunkerton%
}{%
{\protect \APACyear {1997}}%
}]{%
dunkerton1997role}
\APACinsertmetastar {%
dunkerton1997role}%
\begin{APACrefauthors}%
Dunkerton, T\BPBI J.%
\end{APACrefauthors}%
\unskip\
\newblock
\APACrefYearMonthDay{1997}{}{}.
\newblock
{\BBOQ}\APACrefatitle {The role of gravity waves in the quasi-biennial
  oscillation} {The role of gravity waves in the quasi-biennial
  oscillation}.{\BBCQ}
\newblock
\APACjournalVolNumPages{Journal of Geophysical Research:
  Atmospheres}{102}{D22}{26053--26076}.
\PrintBackRefs{\CurrentBib}

\bibitem [\protect \citeauthoryear {%
ECMWF%
}{%
ECMWF%
}{%
{\protect \APACyear {2018}}%
}]{%
ifscy45r1}
\APACinsertmetastar {%
ifscy45r1}%
\begin{APACrefauthors}%
ECMWF.%
\end{APACrefauthors}%
\unskip\
\newblock
\APACrefYearMonthDay{2018}{}{}.
\newblock
\APACrefbtitle {IFS documentation. CY45R1.} {Ifs documentation. cy45r1.}
\newblock
\APAChowpublished
  {\url{https://www.ecmwf.int/en/publications/ifs-documentation}}.
\PrintBackRefs{\CurrentBib}

\bibitem [\protect \citeauthoryear {%
Ern%
, Preusse%
, Alexander%
\BCBL {}\ \BBA {} Warner%
}{%
Ern%
\ \protect \BOthers {.}}{%
{\protect \APACyear {2004}}%
}]{%
ern2004absolute}
\APACinsertmetastar {%
ern2004absolute}%
\begin{APACrefauthors}%
Ern, M.%
, Preusse, P.%
, Alexander, M\BPBI J.%
\BCBL {}\ \BBA {} Warner, C\BPBI D.%
\end{APACrefauthors}%
\unskip\
\newblock
\APACrefYearMonthDay{2004}{}{}.
\newblock
{\BBOQ}\APACrefatitle {Absolute values of gravity wave momentum flux derived
  from satellite data} {Absolute values of gravity wave momentum flux derived
  from satellite data}.{\BBCQ}
\newblock
\APACjournalVolNumPages{Journal of Geophysical Research:
  Atmospheres}{109}{D20}{}.
\PrintBackRefs{\CurrentBib}

\bibitem [\protect \citeauthoryear {%
Garcia%
\ \BBA {} Boville%
}{%
Garcia%
\ \BBA {} Boville%
}{%
{\protect \APACyear {1994}}%
}]{%
garcia1994downward}
\APACinsertmetastar {%
garcia1994downward}%
\begin{APACrefauthors}%
Garcia, R\BPBI R.%
\BCBT {}\ \BBA {} Boville, B\BPBI A.%
\end{APACrefauthors}%
\unskip\
\newblock
\APACrefYearMonthDay{1994}{}{}.
\newblock
{\BBOQ}\APACrefatitle {“Downward control” of the mean meridional
  circulation and temperature distribution of the polar winter stratosphere}
  {“downward control” of the mean meridional circulation and temperature
  distribution of the polar winter stratosphere}.{\BBCQ}
\newblock
\APACjournalVolNumPages{Journal of the atmospheric
  sciences}{51}{15}{2238--2245}.
\PrintBackRefs{\CurrentBib}

\bibitem [\protect \citeauthoryear {%
Gardner%
, Miller%
\BCBL {}\ \BBA {} Liu%
}{%
Gardner%
\ \protect \BOthers {.}}{%
{\protect \APACyear {1989}}%
}]{%
gardner1989rayleigh}
\APACinsertmetastar {%
gardner1989rayleigh}%
\begin{APACrefauthors}%
Gardner, C\BPBI S.%
, Miller, M\BPBI S.%
\BCBL {}\ \BBA {} Liu, C\BHBI H.%
\end{APACrefauthors}%
\unskip\
\newblock
\APACrefYearMonthDay{1989}{}{}.
\newblock
{\BBOQ}\APACrefatitle {Rayleigh lidar observations of gravity wave activity in
  the upper stratosphere at Urbana, Illinois} {Rayleigh lidar observations of
  gravity wave activity in the upper stratosphere at urbana, illinois}.{\BBCQ}
\newblock
\APACjournalVolNumPages{Journal of the atmospheric
  sciences}{46}{12}{1838--1854}.
\PrintBackRefs{\CurrentBib}

\bibitem [\protect \citeauthoryear {%
Geer%
}{%
Geer%
}{%
{\protect \APACyear {2016}}%
}]{%
geer2016significance}
\APACinsertmetastar {%
geer2016significance}%
\begin{APACrefauthors}%
Geer, A\BPBI J.%
\end{APACrefauthors}%
\unskip\
\newblock
\APACrefYearMonthDay{2016}{}{}.
\newblock
{\BBOQ}\APACrefatitle {Significance of changes in medium-range forecast scores}
  {Significance of changes in medium-range forecast scores}.{\BBCQ}
\newblock
\APACjournalVolNumPages{Tellus A: Dynamic Meteorology and
  Oceanography}{68}{1}{30229}.
\PrintBackRefs{\CurrentBib}

\bibitem [\protect \citeauthoryear {%
Gentine%
, Pritchard%
, Rasp%
, Reinaudi%
\BCBL {}\ \BBA {} Yacalis%
}{%
Gentine%
\ \protect \BOthers {.}}{%
{\protect \APACyear {2018}}%
}]{%
gentine2018could}
\APACinsertmetastar {%
gentine2018could}%
\begin{APACrefauthors}%
Gentine, P.%
, Pritchard, M.%
, Rasp, S.%
, Reinaudi, G.%
\BCBL {}\ \BBA {} Yacalis, G.%
\end{APACrefauthors}%
\unskip\
\newblock
\APACrefYearMonthDay{2018}{}{}.
\newblock
{\BBOQ}\APACrefatitle {Could machine learning break the convection
  parameterization deadlock?} {Could machine learning break the convection
  parameterization deadlock?}{\BBCQ}
\newblock
\APACjournalVolNumPages{Geophysical Research Letters}{45}{11}{5742--5751}.
\PrintBackRefs{\CurrentBib}

\bibitem [\protect \citeauthoryear {%
Gettelman%
\ \protect \BOthers {.}}{%
Gettelman%
\ \protect \BOthers {.}}{%
{\protect \APACyear {2020}}%
}]{%
gettelman2020machine}
\APACinsertmetastar {%
gettelman2020machine}%
\begin{APACrefauthors}%
Gettelman, A.%
, Gagne, D\BPBI J.%
, Chen, C\BHBI C.%
, Christensen, M.%
, Lebo, Z.%
, Morrison, H.%
\BCBL {}\ \BBA {} Gantos, G.%
\end{APACrefauthors}%
\unskip\
\newblock
\APACrefYearMonthDay{2020}{}{}.
\newblock
{\BBOQ}\APACrefatitle {Machine learning the warm rain process} {Machine
  learning the warm rain process}.{\BBCQ}
\newblock
\APACjournalVolNumPages{Journal of Advances in Modeling Earth
  Systems}{}{}{e2020MS002268}.
\PrintBackRefs{\CurrentBib}

\bibitem [\protect \citeauthoryear {%
Hatfield%
, Chantry%
, D{\"u}ben%
\BCBL {}\ \BBA {} Palmer%
}{%
Hatfield%
\ \protect \BOthers {.}}{%
{\protect \APACyear {2019}}%
}]{%
hatfield2019accelerating}
\APACinsertmetastar {%
hatfield2019accelerating}%
\begin{APACrefauthors}%
Hatfield, S.%
, Chantry, M.%
, D{\"u}ben, P.%
\BCBL {}\ \BBA {} Palmer, T.%
\end{APACrefauthors}%
\unskip\
\newblock
\APACrefYearMonthDay{2019}{}{}.
\newblock
{\BBOQ}\APACrefatitle {Accelerating high-resolution weather models with
  deep-learning hardware} {Accelerating high-resolution weather models with
  deep-learning hardware}.{\BBCQ}
\newblock
\BIn{} \APACrefbtitle {Proceedings of the Platform for Advanced Scientific
  Computing Conference} {Proceedings of the platform for advanced scientific
  computing conference}\ (\BPGS\ 1--11).
\PrintBackRefs{\CurrentBib}

\bibitem [\protect \citeauthoryear {%
Hatfield%
\ \protect \BOthers {.}}{%
Hatfield%
\ \protect \BOthers {.}}{%
{\protect \APACyear {2021}}%
}]{%
hatfieldTL}
\APACinsertmetastar {%
hatfieldTL}%
\begin{APACrefauthors}%
Hatfield, S.%
, D{\"u}ben, P.%
, Lopez, P.%
, Geer, A.%
, Chantry, M.%
\BCBL {}\ \BBA {} Palmer, T.%
\end{APACrefauthors}%
\unskip\
\newblock
\APACrefYearMonthDay{2021}{}{}.
\newblock
{\BBOQ}\APACrefatitle {{Neural networks as the building blocks for
  tangent-linear and adjoint models}} {{Neural networks as the building blocks
  for tangent-linear and adjoint models}}.{\BBCQ}
\newblock
\APACjournalVolNumPages{arXiv preprint arXiv:}{}{}{}.
\PrintBackRefs{\CurrentBib}

\bibitem [\protect \citeauthoryear {%
Krasnopolsky%
}{%
Krasnopolsky%
}{%
{\protect \APACyear {1997}}%
}]{%
krasnopolsky1997neural}
\APACinsertmetastar {%
krasnopolsky1997neural}%
\begin{APACrefauthors}%
Krasnopolsky, V.%
\end{APACrefauthors}%
\unskip\
\newblock
\APACrefYearMonthDay{1997}{}{}.
\newblock
{\BBOQ}\APACrefatitle {A neural network forward model for direct assimilation
  of SSM/I brightness temperatures into atmospheric models} {A neural network
  forward model for direct assimilation of ssm/i brightness temperatures into
  atmospheric models}.{\BBCQ}
\newblock
\APACjournalVolNumPages{Research activities in atmospheric and oceanic
  modeling}{}{}{}.
\PrintBackRefs{\CurrentBib}

\bibitem [\protect \citeauthoryear {%
Morcrette%
, Mozdzynski%
\BCBL {}\ \BBA {} Leutbecher%
}{%
Morcrette%
\ \protect \BOthers {.}}{%
{\protect \APACyear {2008}}%
}]{%
morcrette2008reduced}
\APACinsertmetastar {%
morcrette2008reduced}%
\begin{APACrefauthors}%
Morcrette, J\BHBI J.%
, Mozdzynski, G.%
\BCBL {}\ \BBA {} Leutbecher, M.%
\end{APACrefauthors}%
\unskip\
\newblock
\APACrefYearMonthDay{2008}{}{}.
\newblock
{\BBOQ}\APACrefatitle {A reduced radiation grid for the ECMWF Integrated
  Forecasting System} {A reduced radiation grid for the ecmwf integrated
  forecasting system}.{\BBCQ}
\newblock
\APACjournalVolNumPages{Monthly weather review}{136}{12}{4760--4772}.
\PrintBackRefs{\CurrentBib}

\bibitem [\protect \citeauthoryear {%
{NVIDIA}%
}{%
{NVIDIA}%
}{%
{\protect \APACyear {2017}}%
}]{%
nvidia2017}
\APACinsertmetastar {%
nvidia2017}%
\begin{APACrefauthors}%
{NVIDIA}.%
\end{APACrefauthors}%
\unskip\
\newblock
\APACrefYearMonthDay{2017}{}{}.
\newblock
\APACrefbtitle {{NVIDIA Tesla V100 GPU Architecture}} {{NVIDIA Tesla V100 GPU
  Architecture}}\ \APACbVolEdTR{}{\BTR{}}.
\newblock
\begin{APACrefURL}
  \url{http://www.nvidia.com/content/gated-pdfs/Volta-Architecture-Whitepaper-v1.1.pdf}
  \end{APACrefURL}
\PrintBackRefs{\CurrentBib}

\bibitem [\protect \citeauthoryear {%
O'Gorman%
\ \BBA {} Dwyer%
}{%
O'Gorman%
\ \BBA {} Dwyer%
}{%
{\protect \APACyear {2018}}%
}]{%
o2018using}
\APACinsertmetastar {%
o2018using}%
\begin{APACrefauthors}%
O'Gorman, P\BPBI A.%
\BCBT {}\ \BBA {} Dwyer, J\BPBI G.%
\end{APACrefauthors}%
\unskip\
\newblock
\APACrefYearMonthDay{2018}{}{}.
\newblock
{\BBOQ}\APACrefatitle {Using machine learning to parameterize moist convection:
  Potential for modeling of climate, climate change, and extreme events} {Using
  machine learning to parameterize moist convection: Potential for modeling of
  climate, climate change, and extreme events}.{\BBCQ}
\newblock
\APACjournalVolNumPages{Journal of Advances in Modeling Earth
  Systems}{10}{10}{2548--2563}.
\PrintBackRefs{\CurrentBib}

\bibitem [\protect \citeauthoryear {%
Orr%
, Bechtold%
, Scinocca%
, Ern%
\BCBL {}\ \BBA {} Janiskova%
}{%
Orr%
\ \protect \BOthers {.}}{%
{\protect \APACyear {2010}}%
}]{%
orr2010improved}
\APACinsertmetastar {%
orr2010improved}%
\begin{APACrefauthors}%
Orr, A.%
, Bechtold, P.%
, Scinocca, J.%
, Ern, M.%
\BCBL {}\ \BBA {} Janiskova, M.%
\end{APACrefauthors}%
\unskip\
\newblock
\APACrefYearMonthDay{2010}{}{}.
\newblock
{\BBOQ}\APACrefatitle {Improved middle atmosphere climate and forecasts in the
  ECMWF model through a nonorographic gravity wave drag parameterization}
  {Improved middle atmosphere climate and forecasts in the ecmwf model through
  a nonorographic gravity wave drag parameterization}.{\BBCQ}
\newblock
\APACjournalVolNumPages{Journal of climate}{23}{22}{5905--5926}.
\PrintBackRefs{\CurrentBib}

\bibitem [\protect \citeauthoryear {%
Ott%
\ \protect \BOthers {.}}{%
Ott%
\ \protect \BOthers {.}}{%
{\protect \APACyear {2020}}%
}]{%
fortrankeras}
\APACinsertmetastar {%
fortrankeras}%
\begin{APACrefauthors}%
Ott, J.%
, Pritchard, M.%
, Best, N.%
, Linstead, E.%
, Curcic, M.%
\BCBL {}\ \BBA {} Baldi, P.%
\end{APACrefauthors}%
\unskip\
\newblock
\APACrefYearMonthDay{2020}{}{}.
\newblock
{\BBOQ}\APACrefatitle {A Fortran-Keras Deep Learning Bridge for Scientific
  Computing} {A fortran-keras deep learning bridge for scientific
  computing}.{\BBCQ}
\newblock
\APACjournalVolNumPages{arXiv preprint arXiv:2004.10652}{}{}{}.
\PrintBackRefs{\CurrentBib}

\bibitem [\protect \citeauthoryear {%
Palmer%
}{%
Palmer%
}{%
{\protect \APACyear {2020}}%
}]{%
palmer2020vision}
\APACinsertmetastar {%
palmer2020vision}%
\begin{APACrefauthors}%
Palmer, T.%
\end{APACrefauthors}%
\unskip\
\newblock
\APACrefYearMonthDay{2020}{}{}.
\newblock
{\BBOQ}\APACrefatitle {A Vision for Numerical Weather Prediction in 2030} {A
  vision for numerical weather prediction in 2030}.{\BBCQ}
\newblock
\APACjournalVolNumPages{arXiv preprint arXiv:2007.04830}{}{}{}.
\PrintBackRefs{\CurrentBib}

\bibitem [\protect \citeauthoryear {%
Polichtchouk%
, Shepherd%
, Hogan%
\BCBL {}\ \BBA {} Bechtold%
}{%
Polichtchouk%
, Shepherd%
, Hogan%
\BCBL {}\ \BBA {} Bechtold%
}{%
{\protect \APACyear {2018}}%
}]{%
polichtchouk2018sensitivity}
\APACinsertmetastar {%
polichtchouk2018sensitivity}%
\begin{APACrefauthors}%
Polichtchouk, I.%
, Shepherd, T.%
, Hogan, R.%
\BCBL {}\ \BBA {} Bechtold, P.%
\end{APACrefauthors}%
\unskip\
\newblock
\APACrefYearMonthDay{2018}{}{}.
\newblock
{\BBOQ}\APACrefatitle {Sensitivity of the Brewer--Dobson circulation and polar
  vortex variability to parameterized nonorographic gravity wave drag in a
  high-resolution atmospheric model} {Sensitivity of the brewer--dobson
  circulation and polar vortex variability to parameterized nonorographic
  gravity wave drag in a high-resolution atmospheric model}.{\BBCQ}
\newblock
\APACjournalVolNumPages{Journal of the Atmospheric
  Sciences}{75}{5}{1525--1543}.
\PrintBackRefs{\CurrentBib}

\bibitem [\protect \citeauthoryear {%
Polichtchouk%
, Shepherd%
\BCBL {}\ \BBA {} Byrne%
}{%
Polichtchouk%
, Shepherd%
\BCBL {}\ \BBA {} Byrne%
}{%
{\protect \APACyear {2018}}%
}]{%
polichtchouk2018impact}
\APACinsertmetastar {%
polichtchouk2018impact}%
\begin{APACrefauthors}%
Polichtchouk, I.%
, Shepherd, T\BPBI G.%
\BCBL {}\ \BBA {} Byrne, N\BPBI J.%
\end{APACrefauthors}%
\unskip\
\newblock
\APACrefYearMonthDay{2018}{}{}.
\newblock
{\BBOQ}\APACrefatitle {Impact of Parametrized Nonorographic Gravity Wave Drag
  on Stratosphere-Troposphere Coupling in the Northern and Southern
  Hemispheres} {Impact of parametrized nonorographic gravity wave drag on
  stratosphere-troposphere coupling in the northern and southern
  hemispheres}.{\BBCQ}
\newblock
\APACjournalVolNumPages{Geophysical Research Letters}{45}{16}{8612--8618}.
\PrintBackRefs{\CurrentBib}

\bibitem [\protect \citeauthoryear {%
Ramachandran%
, Zoph%
\BCBL {}\ \BBA {} Le%
}{%
Ramachandran%
\ \protect \BOthers {.}}{%
{\protect \APACyear {2017}}%
}]{%
swish}
\APACinsertmetastar {%
swish}%
\begin{APACrefauthors}%
Ramachandran, P.%
, Zoph, B.%
\BCBL {}\ \BBA {} Le, Q\BPBI V.%
\end{APACrefauthors}%
\unskip\
\newblock
\APACrefYearMonthDay{2017}{}{}.
\newblock
{\BBOQ}\APACrefatitle {Swish: a self-gated activation function} {Swish: a
  self-gated activation function}.{\BBCQ}
\newblock
\APACjournalVolNumPages{arXiv preprint arXiv:1710.05941}{7}{}{}.
\PrintBackRefs{\CurrentBib}

\bibitem [\protect \citeauthoryear {%
Rasp%
\ \protect \BOthers {.}}{%
Rasp%
\ \protect \BOthers {.}}{%
{\protect \APACyear {2020}}%
}]{%
weatherbench}
\APACinsertmetastar {%
weatherbench}%
\begin{APACrefauthors}%
Rasp, S.%
, D{\"u}ben, P\BPBI D.%
, Scher, S.%
, Weyn, J\BPBI A.%
, Mouatadid, S.%
\BCBL {}\ \BBA {} Thuerey, N.%
\end{APACrefauthors}%
\unskip\
\newblock
\APACrefYearMonthDay{2020}{}{}.
\newblock
{\BBOQ}\APACrefatitle {WeatherBench: A benchmark dataset for data-driven
  weather forecasting} {Weatherbench: A benchmark dataset for data-driven
  weather forecasting}.{\BBCQ}
\newblock
\APACjournalVolNumPages{arXiv preprint arXiv:2002.00469}{}{}{}.
\PrintBackRefs{\CurrentBib}

\bibitem [\protect \citeauthoryear {%
Rasp%
, Pritchard%
\BCBL {}\ \BBA {} Gentine%
}{%
Rasp%
\ \protect \BOthers {.}}{%
{\protect \APACyear {2018}}%
}]{%
rasp2018deep}
\APACinsertmetastar {%
rasp2018deep}%
\begin{APACrefauthors}%
Rasp, S.%
, Pritchard, M\BPBI S.%
\BCBL {}\ \BBA {} Gentine, P.%
\end{APACrefauthors}%
\unskip\
\newblock
\APACrefYearMonthDay{2018}{}{}.
\newblock
{\BBOQ}\APACrefatitle {Deep learning to represent subgrid processes in climate
  models} {Deep learning to represent subgrid processes in climate
  models}.{\BBCQ}
\newblock
\APACjournalVolNumPages{Proceedings of the National Academy of
  Sciences}{115}{39}{9684--9689}.
\PrintBackRefs{\CurrentBib}

\bibitem [\protect \citeauthoryear {%
Rasp%
\ \BBA {} Thuerey%
}{%
Rasp%
\ \BBA {} Thuerey%
}{%
{\protect \APACyear {2021}}%
}]{%
rasp2021data}
\APACinsertmetastar {%
rasp2021data}%
\begin{APACrefauthors}%
Rasp, S.%
\BCBT {}\ \BBA {} Thuerey, N.%
\end{APACrefauthors}%
\unskip\
\newblock
\APACrefYearMonthDay{2021}{}{}.
\newblock
{\BBOQ}\APACrefatitle {Data-driven medium-range weather prediction with a
  Resnet pretrained on climate simulations: A new model for WeatherBench}
  {Data-driven medium-range weather prediction with a resnet pretrained on
  climate simulations: A new model for weatherbench}.{\BBCQ}
\newblock
\APACjournalVolNumPages{Journal of Advances in Modeling Earth
  Systems}{}{}{e2020MS002405}.
\PrintBackRefs{\CurrentBib}

\bibitem [\protect \citeauthoryear {%
Saffin%
, Hatfield%
, D{\"u}ben%
\BCBL {}\ \BBA {} Palmer%
}{%
Saffin%
\ \protect \BOthers {.}}{%
{\protect \APACyear {2020}}%
}]{%
saffin2020reduced}
\APACinsertmetastar {%
saffin2020reduced}%
\begin{APACrefauthors}%
Saffin, L.%
, Hatfield, S.%
, D{\"u}ben, P.%
\BCBL {}\ \BBA {} Palmer, T.%
\end{APACrefauthors}%
\unskip\
\newblock
\APACrefYearMonthDay{2020}{}{}.
\newblock
{\BBOQ}\APACrefatitle {Reduced-precision parametrization: lessons from an
  intermediate-complexity atmospheric model} {Reduced-precision
  parametrization: lessons from an intermediate-complexity atmospheric
  model}.{\BBCQ}
\newblock
\APACjournalVolNumPages{Quarterly Journal of the Royal Meteorological
  Society}{146}{729}{1590--1607}.
\PrintBackRefs{\CurrentBib}

\bibitem [\protect \citeauthoryear {%
Scinocca%
}{%
Scinocca%
}{%
{\protect \APACyear {2003}}%
}]{%
scinocca2003accurate}
\APACinsertmetastar {%
scinocca2003accurate}%
\begin{APACrefauthors}%
Scinocca, J\BPBI F.%
\end{APACrefauthors}%
\unskip\
\newblock
\APACrefYearMonthDay{2003}{}{}.
\newblock
{\BBOQ}\APACrefatitle {An accurate spectral nonorographic gravity wave drag
  parameterization for general circulation models} {An accurate spectral
  nonorographic gravity wave drag parameterization for general circulation
  models}.{\BBCQ}
\newblock
\APACjournalVolNumPages{Journal of the atmospheric sciences}{60}{4}{667--682}.
\PrintBackRefs{\CurrentBib}

\bibitem [\protect \citeauthoryear {%
S{\o}nderby%
\ \protect \BOthers {.}}{%
S{\o}nderby%
\ \protect \BOthers {.}}{%
{\protect \APACyear {2020}}%
}]{%
metnet}
\APACinsertmetastar {%
metnet}%
\begin{APACrefauthors}%
S{\o}nderby, C\BPBI K.%
, Espeholt, L.%
, Heek, J.%
, Dehghani, M.%
, Oliver, A.%
, Salimans, T.%
\BDBL {}Kalchbrenner, N.%
\end{APACrefauthors}%
\unskip\
\newblock
\APACrefYearMonthDay{2020}{}{}.
\newblock
{\BBOQ}\APACrefatitle {MetNet: A Neural Weather Model for Precipitation
  Forecasting} {Metnet: A neural weather model for precipitation
  forecasting}.{\BBCQ}
\newblock
\APACjournalVolNumPages{arXiv preprint arXiv:2003.12140}{}{}{}.
\PrintBackRefs{\CurrentBib}

\bibitem [\protect \citeauthoryear {%
Ukkonen%
, Pincus%
, Hogan%
, Pagh~Nielsen%
\BCBL {}\ \BBA {} Kaas%
}{%
Ukkonen%
\ \protect \BOthers {.}}{%
{\protect \APACyear {2020}}%
}]{%
ukkonen2020accelerating}
\APACinsertmetastar {%
ukkonen2020accelerating}%
\begin{APACrefauthors}%
Ukkonen, P.%
, Pincus, R.%
, Hogan, R\BPBI J.%
, Pagh~Nielsen, K.%
\BCBL {}\ \BBA {} Kaas, E.%
\end{APACrefauthors}%
\unskip\
\newblock
\APACrefYearMonthDay{2020}{}{}.
\newblock
{\BBOQ}\APACrefatitle {Accelerating radiation computations for dynamical models
  with targeted machine learning and code optimization} {Accelerating radiation
  computations for dynamical models with targeted machine learning and code
  optimization}.{\BBCQ}
\newblock
\APACjournalVolNumPages{Journal of Advances in Modeling Earth
  Systems}{12}{12}{e2020MS002226}.
\PrintBackRefs{\CurrentBib}

\bibitem [\protect \citeauthoryear {%
V{\'a}{\v{n}}a%
\ \protect \BOthers {.}}{%
V{\'a}{\v{n}}a%
\ \protect \BOthers {.}}{%
{\protect \APACyear {2017}}%
}]{%
vavna2017single}
\APACinsertmetastar {%
vavna2017single}%
\begin{APACrefauthors}%
V{\'a}{\v{n}}a, F.%
, D{\"u}ben, P.%
, Lang, S.%
, Palmer, T.%
, Leutbecher, M.%
, Salmond, D.%
\BCBL {}\ \BBA {} Carver, G.%
\end{APACrefauthors}%
\unskip\
\newblock
\APACrefYearMonthDay{2017}{}{}.
\newblock
{\BBOQ}\APACrefatitle {Single precision in weather forecasting models: An
  evaluation with the IFS} {Single precision in weather forecasting models: An
  evaluation with the ifs}.{\BBCQ}
\newblock
\APACjournalVolNumPages{Monthly Weather Review}{145}{2}{495--502}.
\PrintBackRefs{\CurrentBib}

\bibitem [\protect \citeauthoryear {%
Veerman%
\ \BBA {} Pincus%
}{%
Veerman%
\ \BBA {} Pincus%
}{%
{\protect \APACyear {2021}}%
}]{%
venmoRad}
\APACinsertmetastar {%
venmoRad}%
\begin{APACrefauthors}%
Veerman, M.%
\BCBT {}\ \BBA {} Pincus, R.%
\end{APACrefauthors}%
\unskip\
\newblock
\APACrefYearMonthDay{2021}{}{}.
\newblock
{\BBOQ}\APACrefatitle {Predicting atmospheric optical properties for radiative
  transfer computations using neural networks} {Predicting atmospheric optical
  properties for radiative transfer computations using neural networks}.{\BBCQ}
\newblock
\APACjournalVolNumPages{Philosophical Transactions A}{}{}{}.
\PrintBackRefs{\CurrentBib}

\bibitem [\protect \citeauthoryear {%
Warner%
\ \BBA {} McIntyre%
}{%
Warner%
\ \BBA {} McIntyre%
}{%
{\protect \APACyear {2001}}%
}]{%
warner2001ultrasimple}
\APACinsertmetastar {%
warner2001ultrasimple}%
\begin{APACrefauthors}%
Warner, C.%
\BCBT {}\ \BBA {} McIntyre, M.%
\end{APACrefauthors}%
\unskip\
\newblock
\APACrefYearMonthDay{2001}{}{}.
\newblock
{\BBOQ}\APACrefatitle {An ultrasimple spectral parameterization for
  nonorographic gravity waves} {An ultrasimple spectral parameterization for
  nonorographic gravity waves}.{\BBCQ}
\newblock
\APACjournalVolNumPages{Journal of the atmospheric
  sciences}{58}{14}{1837--1857}.
\PrintBackRefs{\CurrentBib}

\bibitem [\protect \citeauthoryear {%
Weyn%
, Durran%
\BCBL {}\ \BBA {} Caruana%
}{%
Weyn%
\ \protect \BOthers {.}}{%
{\protect \APACyear {2019}}%
}]{%
weyn2019can}
\APACinsertmetastar {%
weyn2019can}%
\begin{APACrefauthors}%
Weyn, J\BPBI A.%
, Durran, D\BPBI R.%
\BCBL {}\ \BBA {} Caruana, R.%
\end{APACrefauthors}%
\unskip\
\newblock
\APACrefYearMonthDay{2019}{}{}.
\newblock
{\BBOQ}\APACrefatitle {Can machines learn to predict weather? Using deep
  learning to predict gridded 500-hPa geopotential height from historical
  weather data} {Can machines learn to predict weather? using deep learning to
  predict gridded 500-hpa geopotential height from historical weather
  data}.{\BBCQ}
\newblock
\APACjournalVolNumPages{Journal of Advances in Modeling Earth
  Systems}{11}{8}{2680--2693}.
\PrintBackRefs{\CurrentBib}

\bibitem [\protect \citeauthoryear {%
Yuval%
, O'Gorman%
\BCBL {}\ \BBA {} Hill%
}{%
Yuval%
\ \protect \BOthers {.}}{%
{\protect \APACyear {2020}}%
}]{%
yuval2020use}
\APACinsertmetastar {%
yuval2020use}%
\begin{APACrefauthors}%
Yuval, J.%
, O'Gorman, P\BPBI A.%
\BCBL {}\ \BBA {} Hill, C\BPBI N.%
\end{APACrefauthors}%
\unskip\
\newblock
\APACrefYearMonthDay{2020}{}{}.
\newblock
\APACrefbtitle {Use of neural networks for stable, accurate and physically
  consistent parameterization of subgrid atmospheric processes with good
  performance at reduced precision.} {Use of neural networks for stable,
  accurate and physically consistent parameterization of subgrid atmospheric
  processes with good performance at reduced precision.}
\PrintBackRefs{\CurrentBib}

\bibitem [\protect \citeauthoryear {%
Yuval%
\ \BBA {} O’Gorman%
}{%
Yuval%
\ \BBA {} O’Gorman%
}{%
{\protect \APACyear {2020}}%
}]{%
yuval2020stable}
\APACinsertmetastar {%
yuval2020stable}%
\begin{APACrefauthors}%
Yuval, J.%
\BCBT {}\ \BBA {} O’Gorman, P\BPBI A.%
\end{APACrefauthors}%
\unskip\
\newblock
\APACrefYearMonthDay{2020}{}{}.
\newblock
{\BBOQ}\APACrefatitle {Stable machine-learning parameterization of subgrid
  processes for climate modeling at a range of resolutions} {Stable
  machine-learning parameterization of subgrid processes for climate modeling
  at a range of resolutions}.{\BBCQ}
\newblock
\APACjournalVolNumPages{Nature communications}{11}{1}{1--10}.
\PrintBackRefs{\CurrentBib}

\end{thebibliography}
\end{document}